\documentclass[11pt,a4paper]{article}
\usepackage[T1]{fontenc}
\usepackage[utf8]{inputenc}
\usepackage{lmodern}
\usepackage[UKenglish]{babel}
\usepackage[textwidth=150mm,textheight=235mm,centering]{geometry}
\usepackage{amsmath,amssymb}
\usepackage{microtype}
\usepackage[hidelinks]{hyperref}
\usepackage[capitalise,noabbrev]{cleveref}
\usepackage{listings}
\usepackage{xcolor}
\usepackage{booktabs}
\usepackage{graphicx}
\usepackage[bottom]{footmisc}
\usepackage{tikz}
\usetikzlibrary{fit,positioning,decorations.pathreplacing,calc}

\DeclareUnicodeCharacter{03BA}{\ensuremath{\kappa}}
\DeclareUnicodeCharacter{2192}{\ensuremath{\to}}
\lstdefinestyle{lean}{basicstyle=\footnotesize\ttfamily, columns=fullflexible, keepspaces=true,
  frame=single, breaklines=true, mathescape=false, keywords={def,theorem,inductive,where,match,with,fun,structure},
  keywordstyle=\bfseries, comment=[l]{--}, commentstyle=\itshape\color{gray},
  literate={→}{{$\to$}}1 {∀}{{$\forall$}}1 {∃}{{$\exists$}}1 {∧}{{$\wedge$}}1 {∨}{{$\vee$}}1
    {≤}{{$\le$}}1 {¬}{{$\neg$}}1 {∈}{{$\in$}}1 {⟨}{{$\langle$}}1 {⟩}{{$\rangle$}}1
    {₀}{{$_0$}}1 {₁}{{$_1$}}1 {₂}{{$_2$}}1 {κ}{{$\kappa$}}1
    {×}{{$\times$}}1 {←}{{$\leftarrow$}}1 {ₑ}{{$_e$}}1 {σ}{{$\sigma$}}1 {μ}{{$\mu$}}1}
\newcommand{\lean}[1]{\texttt{#1}}

\newcommand{\papertitle}{Order Maintenance in Worst-Case Constant Time, Verified}
\newcommand{\paperauthor}{Denis Firsov}
\newcommand{\paperaffiliation}{Tallinn University of Technology (TalTech)}
\newcommand{\paperemail}{denis.firsov@gmail.com}
\newcommand{\paperkeywords}{order maintenance, worst-case complexity, word RAM, cost semantics, Lean 4}
\newcommand{\paperrepository}{https://github.com/dfirsov/order_maintenance}
\newcommand{\papervisualizer}{https://dfirsov.github.io/order_maintenance/viz/}

\hypersetup{pdftitle={\papertitle},pdfauthor={\paperauthor},pdfkeywords={\paperkeywords}}
\title{\papertitle}
\author{\paperauthor\ifx\paperaffiliation\empty\else\\\paperaffiliation\fi
  \\{\small\href{mailto:\paperemail}{\texttt{\paperemail}}}}
\date{}

\begin{document}
\maketitle
\begin{abstract}
Order maintenance asks for a data structure that keeps a sequence of elements under insertions
and deletions and tells which of two elements comes first. We give an algorithm that performs
each operation in worst-case constant time, and verify it in Lean~4. The
difficulty is to change labels gradually while queries and further insertions continue. The
algorithm, assembled from published components, never relabels the members of a full unit in place:
they move into fresh neighbouring units. Small groups keep element labels short; groups are
collected into larger chunks, which keep accepting insertions while they wait their turn for a
shared top tree; and the top tree never changes a label it has assigned. We implement the algorithm
as a program for a random-access machine (RAM) deeply embedded in Lean, whose interpreter counts every
instruction and allocates a zero-filled array in one step, and
prove that the program answers every query as a simple reference specification does and that
each operation takes at most a fixed number of steps, independent of the number of elements and
of the history. The
machine treats arithmetic on unbounded natural numbers as one step, and on such a machine a
step bound alone is weak, so we also prove that after $t$ operations no value the machine has
computed has more than $O(\log t)$ bits, and that the program runs unchanged, with the same
answers and step counts, on a machine whose words have that many bits. A verified translation
through an intermediate language connects the high-level proofs to the machine program and
preserves step counts exactly, so the interpreter is the only cost model to trust; the development has about 78,000
lines of Lean. Rust code
transpiled from the program agrees with the machine semantics on all tested traces, and an
unverified native implementation shows the algorithm's practical cost.

\end{abstract}

\begingroup\small

\medskip
\noindent\textbf{Supplementary material.} Lean development and transpiled Rust code:\\
\url{\paperrepository}\\
Interactive visualizer: \url{\papervisualizer}
\par\endgroup

\section{Introduction}\label{sec:intro}

The \emph{order-maintenance problem} asks to maintain a sequence under three operations:
insert a new element immediately after a given one, delete an element, and decide whether
one element precedes another. It is a building block of many dynamic algorithms: fully
persistent data structures~\cite{DriscollEtAl89}, self-adjusting
computation~\cite{AcarEtAl06}, race detection for fork-join programs~\cite{BenderEtAl04SP},
and incremental topological ordering~\cite{HaeuplerEtAl12}. The usual approach is
\emph{labeling}: give each element an integer label so that labels increase along the sequence,
and answer a query by comparing two labels. The difficulty is the insertion: repeated
insertions between the same two neighbours eventually use up the labels between them, and
then some labels must change.

The problem goes back to Dietz~\cite{Dietz82}, and constant amortized time per operation is
classical~\cite{Tsakalidis84,DietzSleator87,BenderEtAl02}.
Constant worst-case time has been known since 1987~\cite{DietzSleator87,DietzSleator88}, but
the known algorithms are intricate. Kopelowitz~\cite[Section~1]{Kopelowitz12} summarized the state of the
art: ``to date there is no published version of a worst-case solution for the OMP which is
not considered highly complex''. Bender et al.\ have since simplified worst-case \emph{file
maintenance}~\cite{BenderEtAl17} (keeping items in sorted order in an array with gaps), the component at the top of Dietz and Sleator's construction, but as far as we
know, no worst-case $O(1)$ order-maintenance algorithm has been implemented, and no
order-maintenance algorithm has been verified in a proof assistant.

The obstacle to worst-case constant time is unfinished work: to bound every operation, a
relabeling must be spread over later insertions. Queries then see a partly changed structure, and further insertions
can arrive at the very place being changed, or at other places that also need attention.
A worst-case algorithm must explain both why every intermediate state answers queries
correctly and why the unfinished work never falls behind.

Our construction answers the two questions separately. For correctness, a unit that fills up
is never relabeled in place: it gets two empty neighbours, and its members move into them from
its ends a few at a time, so keys stay ordered at every instant; in addition, every write that
could break the order is checked in constant time. For timing, each insertion pays a fixed
number of steps towards the repairs of the units it makes larger, and a largest-first queue
arbitrates the one place where unrelated units compete. The same scheme serves groups of
elements, chunks of groups, and the nodes of the top tree. \Cref{sec:algorithm}
develops these ideas.

We assemble the algorithm from published components and implement it in a small imperative
language that we deeply embed in Lean~4. The language's interpreter counts every instruction
it executes, so running time is part of the semantics and no part of the computation can
escape the count. We prove two theorems about the program: it answers queries correctly
(Theorem A), and every operation takes at most a fixed number of steps, independent of the
length of the sequence and of the history (Theorem B).

A step bound alone would say little. The machine treats arithmetic on unbounded natural
numbers, and allocation, as single steps, and on such a machine order maintenance is easy:
labels of unbounded length never run out (\cref{sec:background}). What makes the problem hard
is that labels must stay short. We therefore also prove Theorem W: no value the machine
computes, whether stored or intermediate, has more than $O(\log N)$ bits, where $N$ is
the capacity of the structure, except handles and a few wrapper counters, which are bounded by
the number of operations.
The capacity is at most linear in the number of operations, so the $i$-th operation computes
only values of $O(\log i)$ bits, and a machine with words of that size, but otherwise the
same, runs the program with the same answers and step counts. Word size and space thus grow
with the number of operations rather than with the length of the sequence. These, and the machine's one-step allocation of
zero-filled arrays, are the gaps to a textbook word RAM (\cref{sec:limits}).

Checking the literature against our own derivations and simulations, we found errors in two
papers we build on or compare with: the appendix pseudocode of Dietz and
Sleator~\cite{DietzSleator88} has easily repaired slips (two swapped boundary tests, and a
move after the last child of a node that has none), and the proof
of the amortized bound of Bender et al.~\cite[Lemma~3]{BenderEtAl02} miscounts insertions by a
constant factor (the $O(\log n)$ bound survives, but their space/time trade-off does not). A pen-and-paper proof, written before
the formalization, found a cost bug and two gaps in our own design, all of which extensive
testing had missed. The formalization found no further errors, but three parameters had
to change: two because the mechanized arguments are coarser, and one so that an inequality
could be proved for all sizes at once (\cref{sec:errors}).

\paragraph*{Contributions.}
\begin{itemize}
\item \textbf{A complete algorithm designed for proof} (\cref{sec:algorithm}). Its components
  are published: the levels, incremental splitting into fresh neighbours and the analysis of
  largest-first scheduling (the \emph{zeroing game}) come from Dietz and Sleator~\cite{DietzSleator87}, and the weight-balanced tree that replaces
  Willard's algorithm at the top from Kopelowitz~\cite{Kopelowitz12}. Two changes are ours.
  Every write is \emph{guarded} by a constant-time check with a fallback to a linear-time
  relabeling, so the correctness proof of the functional model is local to each write and never
  mentions sizes or timing. The timing proof shows that the fallback never runs; the machine
  program therefore omits it, and its correctness theorem uses both invariants. And the top tree
  splits nodes into fresh neighbours too: the labels of its existing table entries never change, and a split node's
  chunks are redirected to the new nodes in the background (\emph{ghost drains}), so no
  table is ever relabeled. Unlike earlier descriptions, ours leaves no
  case open: every procedure, parameter and constant is fixed, and the proof checks them all.
\item \textbf{Mechanized correctness, a worst-case bound and short words}
  (\cref{sec:formal}): for every valid operation sequence, the machine program returns the
  specification's answers (Theorem A), and every operation takes at most a fixed number of
  machine steps, independent of the length of the sequence and of the history (Theorem B), and computes no value of
  more than $128L + O(1)$ bits, where $L = O(\log N)$, except handles and a few wrapper counters, which are bounded by
  the number of operations (Theorem W); concretely, operation $i$ (counting from $0$) computes no
  value of more than $128\lfloor\log_2(i+1)\rfloor + 1286$ bits. Theorems A and B also hold, for a
  run of $t$ operations, on a machine with a fixed word width of at
  least $128\lfloor\log_2(t+1)\rfloor + 1286$ bits. The development has about $78{,}000$ lines of Lean
  and uses only Lean's standard axioms.
\item \textbf{A machine model in which no computation escapes the count}
  (\cref{sec:spec,app:formal}): a small imperative language deeply embedded in Lean, with a
  total interpreter that counts every step, in the tradition of earlier deeply embedded cost
  semantics (\cref{sec:related}).
\item \textbf{Two executable implementations} (\cref{sec:eval}): Rust code produced from the
  machine program by an unverified transpiler, which matches the Lean semantics step for step
  on all tested traces; and an unverified native implementation, within a small factor of an
  AVL tree on insertions and faster on cold order queries.
\end{itemize}

\section{Background}\label{sec:background}

This section recalls the standard solutions for readers new to the problem, explains why
constant time on average is easy and constant time in the worst case is not, and states the
questions that \cref{sec:algorithm} has to answer.

\paragraph*{The problem.}\label{sec:bg-problem}
An order-maintenance structure holds a sequence of elements and supports three operations.
The operation $\mathit{insertAfter}(x)$ creates a new element immediately after $x$ and returns it;
$\mathit{delete}(x)$ removes $x$; and $\mathit{order}(x,y)$ tells whether $x$ comes strictly
before $y$. Elements are never looked up by value: each operation receives \emph{handles}
returned by earlier insertions, and a valid operation uses only live handles. Starting from
the sequence $[a]$, inserting after $a$ gives an element $b$ and the sequence $[a,b]$; inserting after
$a$ again gives $c$ and the sequence $[a,c,b]$; now $\mathit{order}(c,b)$ is true and
$\mathit{order}(b,c)$ is false. A fourth operation, $\mathit{insertFirst}$, inserts at the front
of the sequence. A balanced search
tree with parent pointers supports all these operations in $\Theta(\log n)$ steps; it is the
AVL tree we compare with in \cref{sec:eval}. The goal is $O(1)$ steps for every operation.

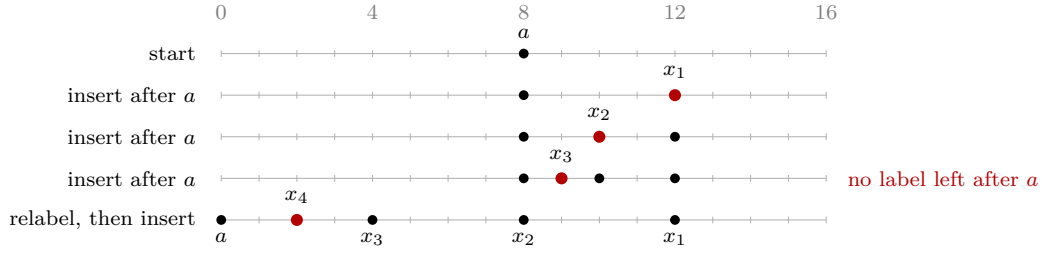
\begin{figure}[t]
\centering
\begin{tikzpicture}[x=0.5cm, y=0.55cm,
  el/.style={circle, fill=black, inner sep=1.3pt},
  new/.style={circle, fill=red!70!black, inner sep=1.6pt},
  lab/.style={font=\scriptsize}]
  \foreach \r/\txt in {0/{start}, 1/{insert after $a$}, 2/{insert after $a$}, 3/{insert after $a$}, 4/{relabel, then insert}} {
    \draw[gray!60] (0,-\r) -- (16,-\r);
    \foreach \t in {0,...,16} { \draw[gray!60] (\t,-\r-0.08) -- (\t,-\r+0.08); }
    \node[lab, anchor=east] at (-0.4,-\r) {\txt};
  }
  \foreach \t in {0,4,8,12,16} { \node[lab, gray] at (\t,1.0) {\t}; }
  \node[el, label={[lab]above:$a$}] at (8,0) {};
  \node[el] at (8,-1) {}; \node[new, label={[lab]above:$x_1$}] at (12,-1) {};
  \node[el] at (8,-2) {}; \node[el] at (12,-2) {}; \node[new, label={[lab]above:$x_2$}] at (10,-2) {};
  \node[el] at (8,-3) {}; \node[el] at (12,-3) {}; \node[el] at (10,-3) {};
  \node[new, label={[lab]above:$x_3$}] at (9,-3) {};
  \node[lab, anchor=west, red!70!black] at (16.3,-3) {no label left after $a$};
  \foreach \t in {0,4,8,12} { \node[el] at (\t,-4) {}; }
  \node[new, label={[lab]above:$x_4$}] at (2,-4) {};
  \node[lab, below=1pt] at (0,-4) {$a$};
  \node[lab, below=1pt] at (4,-4) {$x_3$};
  \node[lab, below=1pt] at (8,-4) {$x_2$};
  \node[lab, below=1pt] at (12,-4) {$x_1$};
\end{tikzpicture}
\caption{Labels in $[0,16)$. Each new element takes the average of the labels around it
(16 bounds the sequence on the right). Three insertions in one place use up the gap after $a$;
the elements are then relabeled evenly, and the fourth insertion fits.}
\label{fig:gaps}
\end{figure}

\paragraph*{Labels with gaps.}\label{sec:bg-labels}\label{sec:bg-ram}
The usual approach gives every element an integer \emph{label}, with labels increasing along
the sequence, so that $\mathit{order}(x,y)$ is a single comparison. A new element takes the
average of its neighbours' labels. Repeated insertions at one place halve the same gap
(\cref{fig:gaps}): between $8$ and $16$, inserting after $8$ takes $12$, then $10$, then $9$,
and then no integer is left. A label of $B$ bits therefore buys only $O(B)$ insertions in one
place before some labels must be spread out again. Choosing which labels to spread out is the
problem of \emph{list labeling}~\cite{ItaiEtAl81}, and a single labeling cannot make this
cheap: when labels come from a polynomial range, an insertion by a deterministic algorithm
needs $\Omega(\log n)$ relabelings on average~\cite{DietzEtAl04,BabkaEtAl12}.
Labels of unbounded length would avoid relabeling altogether, since two binary fractions
always have an average, but after $n$ insertions in one place a label has $n$ bits. Constant
time is meant on a word RAM, where a step operates on a word of $O(\log n)$ bits, so labels
must fit in a constant number of words (\cref{sec:limits} relates this to our machine).

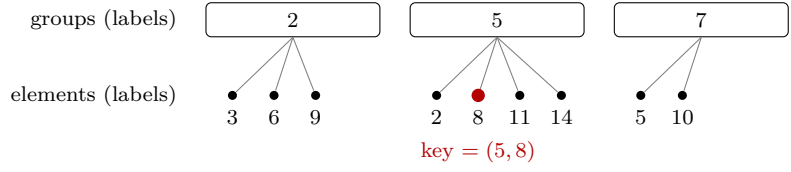
\begin{figure}[t]
\centering
\begin{tikzpicture}[
  el/.style={circle, fill=black, inner sep=1.2pt},
  grp/.style={draw, rounded corners=2pt, minimum height=0.45cm, font=\scriptsize},
  lab/.style={font=\scriptsize}]
  \node[lab, anchor=east] at (-0.4, 1.0) {groups (labels)};
  \node[lab, anchor=east] at (-0.4, 0.0) {elements (labels)};
  \foreach \g/\x/\gl/\labs in {g1/1.0/2/{3,6,9}, g2/3.7/5/{2,8,11,14}, g3/6.4/7/{5,10}} {
    \node[grp, minimum width=2.3cm] (\g) at (\x, 1.0) {\gl};
    \foreach \l [count=\k] in \labs {
      \pgfmathsetmacro\px{\x - 0.8 + 0.55*(\k - 1)}
      \node[el, label={[lab]below:\l}] (\g e\k) at (\px, 0) {};
      \draw[gray] (\g.south) -- (\g e\k);
    }
  }
  \node[el, red!70!black, minimum size=5pt] at (g2e2) {};
  \node[lab, red!70!black] at ($(g2e2)+(0,-0.75)$) {key $= (5, 8)$};
\end{tikzpicture}
\caption{Two-level keys. Group labels order the groups; element labels matter only within a
group, so an insertion relabels at most the elements of its own group.}
\label{fig:twolevel}
\end{figure}

\paragraph*{Indirection: constant time on average.}\label{sec:bg-indirection}
The standard way around this bound is \emph{indirection}~\cite{DietzSleator87,BenderEtAl02}.
The sequence is cut into \emph{groups} of $\Theta(\log n)$ consecutive elements. Each group has a
label in a labeling of the sequence of groups, and each element has a label within its group. The
key of an element is the pair (group label, element label), compared lexicographically
(\cref{fig:twolevel}). Changing a group's label changes the keys of all its elements at once,
through one shared field.

An insertion takes a fresh label within its group, in constant time. A group receives
$O(\log n)$ insertions in its lifetime, so element labels of $O(\log n)$ bits never run out.
When a group becomes too large, it is split in two, both halves are relabeled evenly in
$O(\log n)$ steps, and the new group is inserted into the labeling of groups at an amortized
cost of $O(\log n)$. A group splits only after $\Theta(\log n)$ insertions, so each insertion
costs $O(1)$ on average, and a query still takes two comparisons.

\paragraph*{Why constant worst-case time is hard.}\label{sec:bg-worst}
An amortized bound lets the insertion that triggers a split pay for it. A worst-case bound
does not: a repair must start early and advance by a fixed amount at each later operation.
This raises three questions.
\begin{enumerate}
\item \emph{How can old and new labels coexist?} A group that is half split still receives
  insertions and queries, and keys must increase along the sequence at every instant, not only
  when the split is complete.
\item \emph{Which operations advance a repair, and why does it finish in time?} Insertions
  keep arriving during a split, possibly into the very group being split, and the split must
  end before the groups receiving its elements run out of room.
\item \emph{What happens when repairs compete?} Every split needs a label in the labeling of
  groups. A worst-case labeling, such as Willard's~\cite{Willard92} used by Dietz and
  Sleator, takes $O(\log^2 n)$ steps per insertion, but a new group can appear every
  $\Theta(\log n)$ insertions, so that cost cannot be paid in constant instalments. The remedy
  is more grouping: groups are grouped again into larger units, so that the labeling at the
  top changes rarely enough. Even then, several units may need the top labeling at
  once, and they keep growing while they wait.
\end{enumerate}
\Cref{sec:algorithm} answers them with fresh neighbours, local work budgets, and a
largest-first queue over an extra level of chunks. Changes in the size of the sequence are
handled by global rebuilding~\cite{Overmars83}: a fresh structure is built in the background,
a few elements per operation.

These questions interact. Whether a unit can still accept insertions depends on how long it
waits, which depends on how many units compete, which depends on their sizes. Each mechanism
is simple to state, but the invariants must hold simultaneously after every operation, and
each timing argument assumes that the others held at earlier operations (\cref{sec:why}). This is why the
published descriptions are long, and why constant worst-case time is a natural target for a
mechanized proof. Earlier constructions are
compared with ours in \cref{sec:related,sec:bg-prior}.

\section{The Algorithm}\label{sec:algorithm}

We first explain the layout and a single insertion, then the background work at each level,
and finally why it finishes in time.

\subsection{Overview}\label{sec:overview}

\begin{figure}[b]
\centering
\begin{tikzpicture}[x=0.95cm, y=0.95cm,
  el/.style={circle, fill=black, inner sep=1.1pt},
  grp/.style={draw, rounded corners=2pt, minimum height=0.4cm, minimum width=0.8cm, font=\scriptsize},
  chk/.style={draw=blue!60, very thick, rounded corners=2pt, minimum height=0.42cm, minimum width=1.75cm, font=\scriptsize},
  tn/.style={draw, circle, minimum size=0.42cm, inner sep=0pt, font=\scriptsize},
  lab/.style={font=\scriptsize},
  el2/.style={font=\tiny, inner sep=1pt}]
  \node[lab, anchor=east] at (-0.3, 3.3) {top tree};
  \node[lab, anchor=east] at (-0.3, 1.8) {chunks (tags)};
  \node[lab, anchor=east] at (-0.3, 1.0) {groups (labels)};
  \node[lab, anchor=east] at (-0.3, 0.0) {elements (labels)};
  \foreach \i/\gl in {0/3, 1/9, 2/3, 3/9, 4/3, 5/9, 6/3, 7/9} {
    \node[grp] (g\i) at (0.5 + \i, 1.0) {\gl};
    \foreach \k/\l in {0/2, 1/5, 2/8} {
      \node[el, label={[el2]below:\l}] (e\i\k) at (0.25 + \i + 0.25*\k, 0) {};
      \draw[gray] (g\i.south) -- (e\i\k);
    }
  }
  \foreach \j/\t in {0/0101, 1/0110, 2/1001, 3/1010} {
    \node[chk] (c\j) at (1.0 + 2*\j, 1.8) {\t};
    \pgfmathtruncatemacro\a{2*\j}\pgfmathtruncatemacro\b{2*\j+1}
    \draw (c\j.south) -- (g\a.north); \draw (c\j.south) -- (g\b.north);
  }
  \node[tn] (v0) at (2.0, 2.75) {};
  \node[tn] (v1) at (6.0, 2.75) {};
  \node[tn] (r) at (4.0, 3.5) {};
  \draw (r) -- node[lab, above left=-1pt] {01} (v0);
  \draw (r) -- node[lab, above right=-1pt] {10} (v1);
  \draw (v0) -- node[lab, left] {01} (c0.north);
  \draw (v0) -- node[lab, right] {10} (c1.north);
  \draw (v1) -- node[lab, left] {01} (c2.north);
  \draw (v1) -- node[lab, right] {10} (c3.north);
  \node[el, red!70!black, minimum size=4.5pt] at (e31) {};
  \node[lab, red!70!black] at (3.5, -0.75) {key $= (0110,\ 9,\ 5)$};
  \node[lab, anchor=west, align=left] at (8.3, 3.1) {top jobs, drains\\(\cref{sec:top})};
  \node[lab, anchor=west, align=left] at (8.3, 1.8) {chunk splits, queue\\(\cref{sec:split,sec:reserve})};
  \node[lab, anchor=west, align=left] at (8.3, 1.0) {group splits\\(\cref{sec:split})};
\end{tikzpicture}
\caption{The structure. Groups hold about $L$ consecutive elements and chunks about $L^2$. The
tag of a chunk is the sequence of labels on its path in the top tree. A key is read top-down:
group labels matter only within a chunk, element labels only within a group. On the right,
the background work at each level.}
\label{fig:levels}
\end{figure}
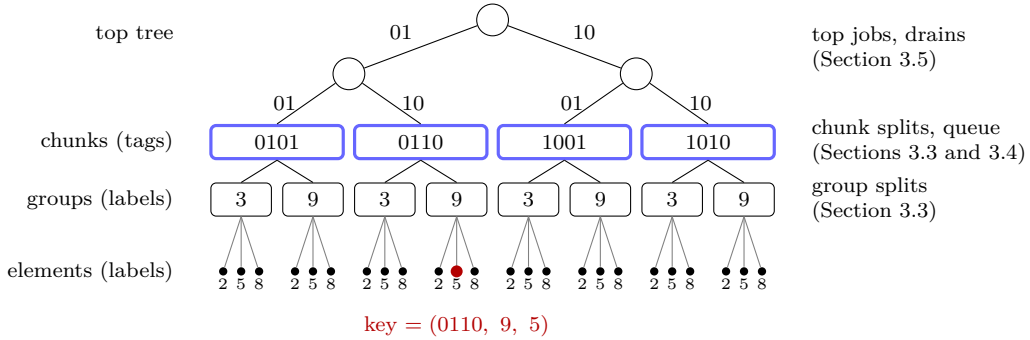

Our structure has four levels (\cref{fig:levels}). Fix a capacity $N \ge 64$, and let its
bit length be $L = \max(8,\lfloor\log_2 N\rfloor + 1)$; all label widths will be $O(L)$.
From the bottom up, the levels are:
\begin{description}
\item[Elements.] The elements of the maintained sequence, in order. Each element carries a \emph{label} that orders it within
  its group.
\item[Groups.] Consecutive elements form \emph{groups} of about $L$ elements. Each group
  carries a label that orders it within its chunk.
\item[Chunks.] Consecutive groups form \emph{chunks} of about $L^2$ elements. Each chunk
  carries a \emph{tag} that orders it among all chunks.
\item[The top tree.] A tree whose leaves are the chunks. The \emph{weight} of a node is the
  number of chunks below it, and a node that becomes too heavy is split, which keeps the tree
  balanced. Each node labels its children, and the tag of a chunk is the sequence of labels on
  its path from the root. The tree together with the jobs that update it is the \emph{top
  structure}.
\end{description}
The \emph{key} of an element is the triple (tag
of its chunk, label of its group, its own label), and an order query compares two keys
lexicographically (see \cref{fig:levels}). An interactive visualizer shows the four levels
while insertions, deletions and rebuilds proceed.\footnote{\url{\papervisualizer}}

We call such a four-level structure, built for a fixed capacity $N$, a \emph{core}. A core
holds at most $N$ elements and never shrinks: a deleted element stays in it as a tombstone.
The data structure the user sees is a \emph{wrapper} that holds one core, or two while it
rebuilds into a core of a different capacity in the background (\cref{sec:wrapper}).

\paragraph*{Why four levels.} Each level makes the level above it change rarely. An insertion
takes a label inside its group, and the groups change only when one splits, about once per
$L$ insertions into it. Group splits in turn change the chunks only when one splits, about
once per $L^2$ insertions into it. Only chunk splits reach the top tree, the one part whose
work grows with the size of the sequence: a \emph{job} there, which enters two new chunks
into the tree (\cref{sec:top}), takes $O(L)$ steps. If groups sat directly
below the top tree, a job would be needed every $\Theta(L)$ insertions, about as fast as jobs
can be paid for, and groups waiting for one would outgrow their labels. Chunks make jobs
rare enough to pay for, with room to wait. Keeping every unit small also keeps every label
short: a unit that receives $O(L)$ new members in its lifetime needs only $O(L)$-bit labels.

\paragraph*{The principle: never relabel in place.} When a unit (a group, a chunk, or a node
of the top tree) approaches its size limit, we do not relabel its members. Two fresh, empty neighbours
are linked on either side of it, and its members migrate into them a few at a time, each
getting a fresh label in its new home. The old unit and its new neighbours are adjacent, so
keys stay sorted at every instant, and a half-finished split is harmless: a query simply
compares the keys as they are. Rebuilding treats a whole core the same way
(\cref{sec:wrapper}).

\paragraph*{The budget: work follows growth.}
\Cref{fig:insert} shows an insertion. It does only one thing for itself: it gives the new
element a label in the group of its predecessor and links it into the element list. Every other line
is its share of the background work, a fixed number of steps at each level
(\cref{tab:params}): it starts a split if its group has just become full, performs $K_G = 4$
moves of the split in its group's \emph{family} (the splitting group and its two new
siblings, \cref{sec:split}) and $K_C = 4$ in its chunk's family, and $K_R = 8$
steps of the top structure, which works on its current job or starts one for the largest
chunk waiting in a queue. The wrapper adds $K_B = 4$ steps of global rebuilding to every
insertion and deletion. A deletion marks its element as a \emph{tombstone}; a query compares
two keys and does no background work. So no operation performs a long repair at once, and the
proof must show that every repair can wait for later insertions. The units grow with the
capacity, so in a larger core the background jobs are longer but proportionally rarer, and
the share per operation stays the same. Every line takes constant time, so the cost proof
must show that the guards never fail and that every table the top structure touches (\cref{sec:top}) stays of
constant size.

\begin{figure}[t]
\begin{lstlisting}[mathescape, basicstyle=\small\ttfamily, columns=fullflexible, frame=single]
insertAfter(x):                       -- in a core
  g := group(x)
  new element y; choose a guarded label for it after x in g
  link y after x;  size(g) += 1;  size(chunk(g)) += 1
  if g just reached G and is not splitting: start its split
  $K_G$ moves of the split in g's family (if any)
  $K_C$ moves of the split in chunk(g)'s family (if any)
  $K_R$ steps of the current top job (or start one from the queue)
  return y
\end{lstlisting}
\caption{An insertion. The wrapper (\cref{sec:wrapper}) adds $K_B = 4$ steps of global
rebuilding to every user update; queries do no background work.}
\label{fig:insert}
\end{figure}

\begin{table}[t]
\centering
\caption{Parameters for a core of capacity $N$, $L = \max(8,\lfloor\log_2 N\rfloor + 1)$.}
\label{tab:params}
\begin{tabular}{lll}
\toprule
Symbol & Meaning & Value \\
\midrule
$G$ & group size at which a split starts & $2L$ \\
$T$, $\theta$ & largest chunk size at split start; queue threshold & $3L^2$, $2L^2$ \\
$K_G, K_C$ & split moves per insertion (groups, chunks) & $4, 4$ \\
$K_R$ & top job steps per insertion & $8$ \\
$a$, $H$ & branching parameter and height of the top tree & $8$, $\lfloor L/3\rfloor + 1$ \\
$s$ & sweep steps per funding job (\cref{sec:top}) & $11$ \\
$\beta$ & label bits per table entry & $64$ \\
$B_e$, $B_g$ & label bits within a group, within a chunk & $O(L)$ (\cref{app:params}) \\
$K_B$ & rebuild budget per update & $4$ \\
\bottomrule
\end{tabular}
\end{table}

The rest of this section explains the lines of \cref{fig:insert}: labels and guards
(\cref{sec:keys}), splits (\cref{sec:split}), the queue of chunks waiting for new tags
(\cref{sec:reserve}), the top structure (\cref{sec:top}), and the rebuilding
(\cref{sec:wrapper}).

\subsection{Labels and guards}\label{sec:keys}\label{sec:guards}

\paragraph*{Labels.} Labels are numbers of a fixed width, chosen for each core: element
labels have $B_e$ bits and lie in $[0, 2^{B_e})$, and group labels have $B_g$ bits and lie in
$[0, 2^{B_g})$. A new label is made in one of three ways: the first member of an empty group
(or chunk) gets the middle of the range; a member added at an end gets its neighbour's label
plus or minus a fixed step $D$; a member added between two others gets the average of their
labels. A label never changes until a split moves its member to another group or chunk
(\cref{sec:split}). Each average halves a gap, so a gap survives only about $\log_2 D$
insertions at the same place. The widths are therefore chosen from the number of new members
a unit receives in its lifetime, which the splits keep at $O(L)$: a group receives $O(L)$
elements, and a chunk, although it receives $O(L^2)$ elements, receives only $O(L)$ groups:
$O(L)$ moved in by the chunk split that created it, and two for each group split, which
happens only after $\Omega(L)$ insertions (\cref{app:params}). All widths are $O(L)$, so a key
fits in a constant number of words, and
reading a key follows the element's group reference and the group's chunk reference. An order
query compares at most three pairs of labels.

\paragraph*{Guards.} Without guards, whether keys stay in the order of the sequence would depend on the timing analysis: if a
group received more insertions than its label width allows, a gap would close. We do not let
correctness rest on that analysis. Every write that could break the order is \emph{guarded}:
before writing, it checks the new value against its neighbours, in constant time. A new label
must lie strictly between its neighbours' labels and inside its range, a new chunk tag must
lie strictly between the tags of the neighbouring chunks, and a split move checks that the
destination sibling is still the unit's immediate neighbour. If a guard fails, a
\emph{fallback} rebuilds the core from its \emph{element list} (all its elements in sequence
order, tombstones included, after a sentinel element) in linear time, with evenly spaced
labels. The sentinel also gives $\mathit{insertFirst}$ its meaning in a core: it inserts after
the sentinel.

This splits the proof in two. The correctness invariant does not need the timing
inequalities: it needs only that the label ranges are wide enough for the fallback's evenly
spaced labels, and every guarded primitive then preserves order or invokes the fallback. Its
reasoning is local to each write, and at the level of the functional implementation it holds
whether or not guards fail (\cref{sec:proof}). The cost invariant shows that no guard ever
fails, so the linear-time fallback never runs. The machine program therefore omits the
fallback, and its correctness theorem uses both invariants. What the guards buy is a
correctness proof that is local to each write and never mentions sizes or timing, not a
correctness theorem that is independent of timing.

\subsection{Splitting groups and chunks}\label{sec:split}

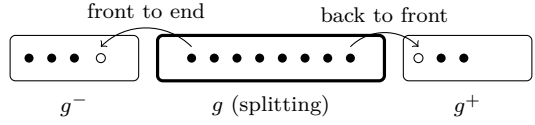
\begin{figure}[t]
\centering
\begin{tikzpicture}[
  el/.style={circle, fill=black, inner sep=1.2pt},
  ghost/.style={circle, draw=black, inner sep=1.2pt},
  box/.style={draw, rounded corners=2pt, minimum height=0.6cm},
  lab/.style={font=\scriptsize}]
  \node[box, minimum width=1.7cm] (gm) at (0,0) {};
  \node[box, minimum width=3.0cm, very thick] (g) at (2.6,0) {};
  \node[box, minimum width=1.7cm] (gp) at (5.2,0) {};
  \node[lab, below=1pt of gm] {$g^-$};
  \node[lab, below=1pt of g] {$g$ (splitting)};
  \node[lab, below=1pt of gp] {$g^+$};
  \foreach \x in {-0.6,-0.3,0.0} { \node[el] at (\x,0) {}; }
  \node[ghost] at (0.35,0) {};
  \foreach \x in {1.55,1.85,...,3.7} { \node[el] at (\x,0) {}; }
  \foreach \x in {4.85,5.15} { \node[el] at (\x,0) {}; }
  \node[ghost] at (4.55,0) {};
  \draw[->, bend right=45] (1.55,0.15) to node[lab, above] {front to end} (0.35,0.15);
  \draw[->, bend left=45] (3.65,0.15) to node[lab, above] {back to front} (4.55,0.15);
\end{tikzpicture}
\caption{An incremental split. Each move takes one member from an end of $g$ into the
adjacent sibling, where it gets a fresh label (hollow circles mark where the moved members
land). Keys stay sorted at every instant because
$g^-$, $g$ and $g^+$ are consecutive.}
\label{fig:split}
\end{figure}

A group that reaches $G = 2L$ elements starts to \emph{split} (\cref{fig:split}): two
empty groups $g^-$ and $g^+$ are linked immediately before and after $g$, and from then on
every insertion into this \emph{family} $(g^-, g, g^+)$ performs $K_G = 4$ \emph{moves}.
A move takes the first element of $g$ to the end of $g^-$, or the last element of $g$ to the
front of $g^+$, whichever sibling is smaller, and gives the element a fresh label there. The
element list itself never changes; only group membership and labels do. When $g$ is empty
it is unlinked and its siblings become ordinary groups.

For example, suppose $g$ contains $[a,b,c,d]$. After moving $a$ left and $d$ right, the
three groups contain $[a]\mid[b,c]\mid[d]$. Their group labels are in the same order,
so a query sees $a<b<c<d$ regardless of the fresh local labels given to $a$ and $d$.
Moving an interior element instead would break this argument: it is essential that moves
take a prefix to the left or a suffix to the right.

Each insertion into the family adds one element and moves four, so $g$ empties after about
$G/3$ insertions. By then the family has gained only about $G/3$ elements, so the two
siblings together hold about $4G/3$, and moving to the smaller sibling keeps them about equal.
So every group stays below a constant times $G$ and receives $O(L)$ insertions in its
lifetime, which is what the label width $B_e$ pays for. If insertions into the family stop,
the split can remain unfinished: nothing there is growing, and the keys already give the
right answers.

Chunks split in the same way, with groups as members and $K_C = 4$ moves per insertion.
The one difference is when a split starts: the new siblings of a chunk need tags from the top
structure, so a large chunk first waits in a queue (\cref{sec:reserve}). Moving a group takes constant time however many elements it has, because elements
point to their group and not to their chunk. One refinement keeps the two levels
independent: if the group about to leave a chunk belongs to a splitting family, the whole
family moves in the same step, so a family never straddles two chunks. Otherwise a later
group move could add elements to another chunk without an insertion there, defeating that
chunk's growth accounting.

\subsection{Waiting for the top: the queue}\label{sec:reserve}

Splitting a chunk creates two new chunks, which must be entered into the top structure
(\cref{sec:top}). Their tags are cheap and are assigned at once. What takes time is the
\emph{job} that records the new chunks in the tree: it updates the weights on the path to the
root and rebalances where needed. A job takes $O(L)$ steps, every insertion advances it by
$K_R = 8$ steps, and only one runs at a time, so that the weights it reads are exact. Chunks
therefore have to wait their turn, and the largest is served first, as in the splitting lemma of
Levcopoulos and Overmars~\cite{LevcopoulosOvermars88} and the reservation process of Dietz and
Sleator~\cite{DietzSleator87}. A bucket queue, as in Kopelowitz's
construction~\cite{Kopelowitz12}, holds the chunks larger than
$\theta = 2L^2$, keyed by size: one bucket per size, with the non-empty buckets linked in
order. A chunk's size changes by one at a time, so it moves to a neighbouring bucket in
constant time; the general insertion into the queue scans for the next non-empty bucket, and
the cost proof shows that this linear-time path is never taken. Whenever no job is running, the largest chunk is taken from
the queue, its two empty siblings are linked and given tags, a job for them starts, and the
chunk starts splitting. The new chunks receive elements without waiting for the job to
finish.

How large can a chunk grow while it waits? This matters because the size of a chunk bounds
the insertions its group labels must absorb. Insertions keep arriving while a job runs, and
the user may aim them all at waiting chunks. Dietz and Sleator~\cite{DietzSleator87} answer
this question with a game, which we call the \emph{zeroing game}. There are $m$ counters, all
starting at $0$. In each round, the
player resets a largest counter to $0$, and then an adversary adds a total of at most $R$ to
the counters, split among them as it likes. Then no counter ever exceeds $R \cdot (\mathcal{H}_{m-1} +
1)$, where $\mathcal{H}_k = 1 + \frac12 + \dots + \frac1k \approx \ln k$ is the $k$-th harmonic
number. The bound is reached roughly
when the adversary spreads its additions evenly over the counters still in play: with $k$
counters left, each gets $R/k$ per round while the player removes one, so the last survivor
collects $R/m + R/(m-1) + \dots + R/1$.

In our setting, a counter is the amount by which a chunk exceeds $\theta$ (zero if it does
not), a round lasts from one job's start to the next, and the reset is the start of a job for
the largest chunk. A chunk that is splitting is not in the queue, and its counter is $0$; the local split
budget, not the game, bounds its growth. A chunk that has just finished splitting is below
$\theta$, so it re-enters the game at $0$.
A job takes $O(L)$ steps and each insertion advances it by $K_R = 8$ steps, so a round
contains $R = O(L)$ insertions, each adding one element to one chunk. There are at most
$2^{L-2}$ counters, because a chunk is split only after growing from the size of a freshly
split chunk to more than $\theta$, and the core holds fewer than $2^L$ elements; so
$\mathcal{H}_{m-1} + 1 = O(L)$, and a chunk exceeds $\theta$ by at most $O(L^2)$
before its job starts. The constants are chosen so that this excess is at most $L^2$, so no
waiting chunk exceeds $T = \theta + L^2 = 3L^2$ before service. The local split budget
then bounds its further growth. This is why chunks hold about $L^2$ elements: one factor of
$L$ covers the length of a job, and the other the waiting factor $\mathcal{H}_{m-1} + 1$, which even
largest-first service cannot avoid.

\subsection{The top structure}\label{sec:top}

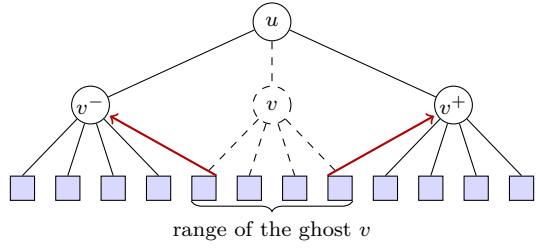
\begin{figure}[t]
\centering
\begin{tikzpicture}[
  node/.style={draw, circle, minimum size=0.5cm, inner sep=0pt, font=\scriptsize},
  ghost/.style={draw, dashed, circle, minimum size=0.5cm, inner sep=0pt, font=\scriptsize},
  ch/.style={draw, fill=blue!15, minimum width=0.32cm, minimum height=0.32cm, inner sep=0pt},
  lab/.style={font=\scriptsize}]
  \node[node] (p) at (3.2,2.2) {$u$};
  \node[node] (vm) at (0.8,1.1) {$v^-$};
  \node[ghost] (v) at (3.2,1.1) {$v$};
  \node[node] (vp) at (5.6,1.1) {$v^+$};
  \draw (p) -- (vm); \draw[dashed] (p) -- (v); \draw (p) -- (vp);
  \foreach \i in {0,...,11} { \node[ch] (c\i) at (0.6*\i - 0.1, 0) {}; }
  \foreach \i in {0,1,2,3} { \draw (vm) -- (c\i.north); }
  \foreach \i in {4,5,6,7} { \draw[dashed] (v) -- (c\i.north); }
  \foreach \i in {8,9,10,11} { \draw (vp) -- (c\i.north); }
  \draw[decorate, decoration={brace, mirror, amplitude=4pt}] (c4.south west) -- (c7.south east)
    node[midway, below=4pt, lab] {range of the ghost $v$};
  \draw[->, thick, red!70!black, shorten >=1pt] (c4.north east) -- (vm);
  \draw[->, thick, red!70!black, shorten >=1pt] (c7.north west) -- (vp);
\end{tikzpicture}
\caption{A ghost drain. The ghost $v$ (dashed) still lies on the stored paths of the chunks in
its range. A sweep step redirects the chunk at an end of the range to $v^-$ or $v^+$ and
rewrites its tag; all table labels stay as they are.}
\label{fig:top}
\end{figure}

The top structure gives every chunk a tag that can be compared directly. The chunks are the leaves of a weight-balanced
tree~\cite{ArgeVitter03}, as in Kopelowitz's construction~\cite{Kopelowitz12}, with branching parameter $a = 8$ (\cref{fig:top}). The chunks have
height $0$, and a node has height $h$ if its children have height $h-1$. Recall that the weight of
a node is the number of chunks below it, not the number of elements. A non-root node of height $h$ is
split when its weight exceeds $2a^h$, which keeps its weight at $O(a^h)$; a node created by a
drain (below) is not split while that drain is unfinished, and \cref{sec:why} shows that it
does not reach the threshold in that time. The height $H = O(L)$
is fixed when the core is created, large enough for the most chunks the core can hold: the
tree widens as chunks are added, but never grows taller. Its root has enough room for
the core's whole lifetime and never needs to split. Every node has a \emph{table}: its
children in order, each with a $\beta$-bit label. A chunk stores its path to the root, and
its tag is the concatenation of the labels along the path, $H\beta = O(L)$ bits in total.
The labels of existing table entries are \emph{immutable}: a new entry gets the average of its neighbours'
labels, and no label ever changes. So a tag changes only when its chunk's path changes.

A job inserts two chunks into the tree and keeps the tree balanced, so that tags stay short
and table labels never run out. A job that inserts two chunks $y, z$ around a chunk $x$ first
gives them tags at once, computed from $x$'s tag (they enter $x$'s lowest-level table right
before and after it), then copies $x$'s stored path to them one level at a time, and finally
walks up the tree from $x$, one node per step, incrementing weights. Only one job runs at a time, so
the weight the walk finds at a node is exact. If it exceeds the threshold, the node $v$ is
split on the spot by the principle of \cref{sec:overview}, here called a \emph{drain}: two new
nodes $v^-, v^+$ are inserted right before and after $v$ in every table that contains $v$, and
$v$ becomes a \emph{ghost}. (A node can sit in more than one table: if its parent is a ghost,
the node stays in the ghost's table until the ghost's range is empty, and also appears in the
table of one of the parent's new nodes.) The live children of $v$ are cut where their weights
reach half of $v$'s weight, but never between the two new nodes of a drain still running below
$v$. Each new node gets a fresh, evenly spaced table of its part. A ghost child whose range is
not yet empty goes to the side that receives its own left new node. The chunks whose stored
paths still pass through the ghost form a contiguous range. A \emph{sweep step} takes the chunk
at the left end of the range if it belongs to $v^-$'s part, and otherwise the chunk at the
right end, and redirects its path to $v^-$ or $v^+$, rewriting its tag; the ghost disappears
when its range is empty.

During a drain the tree already lists $v$'s children under $v^-$ and $v^+$, but the stored
paths of some chunks, and hence their tags, still pass through $v$. Queries read only tags.
Suppose the parent's table gives $v^-$, $v$, $v^+$ the labels $20$, $40$, $60$. Then chunks
already swept left have $20$ in that position of their tags, the unswept range has $40$, and
those swept right have $60$; the positions further down order the chunks within each part. A
sweep step takes a chunk from an end of the $40$-range, so the three parts stay in order, and
it changes only the affected path entry and two tag fields, not the whole path: the field at
height $h$, which holds the label of $v$ in its parent's table and now gets that of $v^-$ or
$v^+$, and the field below it, which holds the label of the next node down in $v$'s table and
now gets its label in the new node's table.

A job whose walk passes $v^-$ or $v^+$ performs $s = 11$ sweep steps of $v$; we call it a
\emph{funding job} of the drain. A new chunk copies its neighbour's stored path, so a
funding job can add at most two chunks to the ghost's range, while it removes eleven: it
reduces the unfinished sweep by at least nine, unless the sweep finishes. Funding jobs are
also the only jobs that make $v^-$ and $v^+$ heavier. These start near half their weight
limit; they gain only two chunks per funding job and cannot reach that limit before the sweep
finishes. If no more such jobs arrive, the drain can remain unfinished safely.

Starting a drain touches only node tables, not all chunks below the node. The weight
invariants bound the number of live children of a node by $O(a)$. Ghost entries are bounded
too, and the argument runs from weights to tables, not the other way round. An entry is added
to a table only by an insertion event (two chunks entering a lowest-level table, or a child
being split), each event requires the weight below the table to grow by a fixed amount, and
that weight is bounded; so the events are bounded, and with them both the length of the table
and the number of times its gaps are halved. For our parameters every table has at most 160
entries, live or ghost, and $\beta = 64$ bits leave room for all its halvings
(\cref{sec:room}). Table operations therefore take a fixed number of steps. The potentially large work is the sweep over chunks, which is spread
over later jobs. Drains at different heights may coexist without waiting for one another;
a job spends at most $s$ sweep steps at each height, so the whole job takes $O(H)=O(L)$
steps. For our parameters the bound proved in Lean is $13H+2$ job steps; each job step is a
bounded number of machine instructions.

Drains are where we depart from earlier work. Willard's algorithm~\cite{Willard92} and
Kopelowitz's construction~\cite{Kopelowitz12} relabel in place, in the background, so the
structure passes through states in which some labels are old and some new. Dietz and Sleator
keep two time-stamped versions of Willard's file, so that queries see a consistent
one~\cite{DietzSleator88}; Kopelowitz rewrites the tags below a split node one at a time, with
an extra most significant bit that orders the rewritten ones after the others. With immutable labels there is no such window: every chunk's tag
is always a valid key, and a sweep step changes one tag, guarded against its two neighbours.
The invariant that becomes simpler is the order of tags, which needs no case for a
half-rewritten set of labels. What drains add is bookkeeping: ghost entries in tables, stored paths
that may pass through ghosts, and the funding jobs that bound how long a drain lasts. The price
is label width, one $\beta$-bit field per level of the tree, and memory for the stored paths
and ghost entries.

\subsection{Deletion and growth}\label{sec:wrapper}

Cores never shrink: a deleted element stays in its core as a \emph{tombstone}. Deletion and
growth are handled by global rebuilding~\cite{Overmars83}. When the current core is half full
(tombstones included), or the sequence has shrunk below an eighth of the capacity, the wrapper
creates a new core. Its capacity is four times the sum of the live-element count and a bound on
insertions arriving during the copy, with a minimum capacity of 64. The second term matters:
even if few elements are live, the copy described below may spend many updates crossing
tombstones while new elements arrive (\cref{sec:errors}). From then on every user update has a budget of $K_B = 4$ rebuild steps. First the new core's
top tree is built, one node per step. Then a cursor walks through the current core's element
list and copies each live element to the end of the new core, skipping tombstones. Meanwhile
operations go to the current core and are mirrored in the new one when they concern the part
already copied: an insertion after a copied element is repeated there, and a deletion of a
copied element is marked there too. When the cursor passes the last element, the new core
has the same live sequence in the same order, and it replaces the old one in a constant number
of steps: the wrapper swaps the roles of its two core slots. The new core may also contain
tombstones of elements deleted after they were copied.

The new core is built from scratch: its capacity, label widths and tree height differ, and its
groups, chunks and labels are all new. The user does not notice, because a handle does not
point into a core. The $k$-th insertion returns the number $k$ as its handle, and the wrapper
keeps, for each core, an array that maps handles to the elements of that core. A query looks
up both handles in the current core's array and compares the keys there. The copy step and
the mirrored insertions fill in the new core's array as they go, and when the new core takes
over, so does its array. So a handle returned long before a rebuild can be compared with one
returned after it: at the time of the query, both are looked up in the same core.

\subsection{Why the budgets suffice}\label{sec:why}

Each piece of background work races the unit it serves: it must finish before that unit runs
out of room. At the group and chunk levels the race is local: the insertions that fill a
split's destinations are the same insertions that drive the split, and each of them moves
four members out while adding at most one. The top tree has the same kind of race. When a
node $v$ is split, its chunks are moved to the two new nodes one at a time by sweep steps, and
the new nodes must not become too heavy, and need splitting themselves, before $v$ is empty.
The only jobs that add chunks below the new nodes are those whose walk passes through them, and
each such job also performs $s = 11$ sweep steps of $v$, so the move finishes in time. The queue is the one place where
unrelated units compete, and there the zeroing game bounds how much a waiting chunk grows.
Global rebuilding follows the same pattern at the scale of a core: it starts while the old
core is only half full, and the new capacity leaves room for the insertions that arrive
during the copy.

The proof turns each race into an inequality between the parameters of \cref{tab:params}.
Seven are needed; \cref{app:budgets} lists them all. Their exact forms, with all rounding,
are proved in Lean for every $L \ge 8$; we show simplified versions. Three of them show the
pattern.
\begin{itemize}
\item \emph{Group splits.} A split lasts at most $k_G = \lfloor G/3\rfloor + 1$ insertions,
  and moving to the smaller sibling leaves each sibling with about half of $G + k_G$
  elements, which must be below the split size:
  \[ \lfloor (G + k_G + 3)/2 \rfloor < G. \]
  So a sibling never needs to split while its own family is still splitting.
\item \emph{Waiting chunks.} A job takes at most $13H + 2$ steps, so a round of the zeroing
  game contains at most $R = \lfloor (13H + 2)/K_R\rfloor + 1$ insertions. With $m$ counters, a
  waiting chunk stays below $T$ if
  \[ \theta + R \cdot (\mathcal{H}_{m-1} + 1) \le T. \]
  With $H = \lfloor L/3\rfloor + 1$ and $K_R = 8$ we get $R \le L$, and $m - 1 < 2^{L-2}$
  gives $\mathcal{H}_{m-1} + 1 \le L$, so the left side is at most $2L^2 + L^2 = 3L^2$.
\item \emph{Drains.} A new node starts with half of its source's threshold $W_h = 2a^h$, plus
  the weight that the cut cannot divide: one child of $v$, or the two new nodes of a drain
  still running below $v$. A child may gain weight during its own drain, so this is up to about
  $W_{h-1} \cdot s/(s-2) = W_h \cdot s/(a(s-2))$. The new node then gains $2$ per funding job for at most
  about $W_h/(s-2)$ jobs. It must stay below the threshold:
  \[ \Bigl(\tfrac12 + \tfrac{s}{a(s-2)} + \tfrac{2}{s-2}\Bigr) \cdot W_h + O(1) \le W_h. \]
  For $a = 8$ and $s = 11$ the coefficient is $7/8$; with $s = 8$ it is $1$, and the
  condition fails.
\end{itemize}
The invariants depend on one another. The top tree's invariants, which bound the length of a
job, need a bound on the number of chunks. That bound holds because a chunk is split only after
growing from the size of a freshly split chunk to more than $\theta$, so it needs the bound $T$
on the size of a chunk when its split starts. That bound comes from the zeroing game, whose
rounds are as long as top jobs. No invariant can therefore be established on its own. They are
proved together, by induction over operations: if all of them hold before an operation, each
one holds after it, using the others as they were before. So no
invariant ever fails; in particular, no guard fails and the fallback never runs.

\section{Formalization}\label{sec:formal}

\subsection{Specification and machine}\label{sec:spec}\label{sec:model}
The specification is a list of distinct element names and a counter for the next fresh name.
$\mathit{insertFirst}$ puts the fresh name at the front of the list, $\mathit{insertAfter}(x)$
right after $x$, $\mathit{delete}$ erases a name, and an order query is answered by
\begin{lstlisting}[style=lean]
def precedes (l : List Elem) (x y : Elem) : Bool := decide (l.idxOf x < l.idxOf y)
\end{lstlisting}
A sequence of operations is \emph{valid} if every argument is live when its operation is
issued; \lean{Spec.run} returns the answers.

We define a random-access machine (RAM) in Lean. Its words are unbounded natural numbers
(\cref{sec:memwords} bounds the values the algorithm stores), its memory is a set of named
arrays with explicit capacities, and its programs are structured: assignments, array loads
and stores, allocation, conditionals, bounded loops, calls, and \lean{fail} and \lean{catch}
(\cref{fig:syntax}). Its interpreter \lean{exec} is a total Lean function that returns the
final state, a status and a step count, charging one step for every expression node,
assignment, memory access, allocation, branch, loop iteration, call, returned value,
\lean{skip} and handler entry. This deep embedding is
deliberate. If the algorithm were ordinary Lean code in a monad that counts designated
primitives, it could do uncounted work between them, such as a list traversal hidden in a
helper function, and ruling that out would be a matter of reviewing code, not of proof. Here
a program has no way to compute outside the interpreter, so a reader only has to check that
\lean{exec} charges every instruction; it and the expression evaluator take 171 lines.

\subsection{The main theorem}\label{sec:main}
\Cref{fig:main} states the main theorem for the machine program. It combines three statements:
the program answers as the specification does (\emph{Theorem A}), every operation takes at
most $\lean{opB}\ K_B$ steps (\emph{Theorem B}), and no operation computes a large value
(\emph{Theorem W}). The initial machine state \lean{m}$_0$ is arbitrary except for the local
\lean{kb}, which holds the rebuild budget $K_B \ge 4$; the hypothesis $K_B < 2^{512}$ only
ensures that $K_B$ itself, which the program stores, is small. The program \lean{initWS}
initializes the structure, and \lean{runMW} then runs the machine procedure for each operation
in turn, returning for each operation its answer, its step count and the largest value it
computes. For every
valid operation sequence the answers are those of \lean{Spec.run}, and every step count is at
most $\lean{opB}\ K_B$, a number that depends on $K_B$ alone: not on $N$, the number of live
elements, or the history. The theorem compares the answers of the queries, which
\lean{ansOf} reads from the machine's local \lean{r}. Handles are not part of the compared answers: in both the specification and the machine
program, the $k$-th insertion returns handle $k$, and the insertion lemma \lean{insertW\_ok}
proves that the machine's result local holds exactly the specification's fresh handle. The budget
$K_B$ is read from a local so that one theorem covers every budget $K_B \ge 4$.

For $K_B = 4$, initialization takes at most $6{,}648$ steps and every operation at most
$1{,}133{,}561{,}742$. The bound is deliberately loose: most of it comes from nested
traversals of bounded-size top-tree tables, multiplied by the fixed number of core insertions
an update can perform (\cref{app:formal}). It shows that a uniform constant exists; it does
not predict running time.

For Theorem W, the interpreter records in an extra field, which no instruction reads, the largest value of every expression
and subexpression it evaluates, including indices and loop bounds (\lean{bump},
\cref{fig:exec}), and \lean{dpk c m} is that value for a run of \lean{c} from \lean{m}.
Initialization computes no value of more than $1{,}029$ bits, and operation $i$ (counting from
$0$) none of more than $128\lfloor\log_2(i+1)\rfloor + 1286$ bits. The bound grows with $i$
because handles are never reused and because the capacity is bounded only in terms of the
number of operations (\cref{sec:limits}).

\paragraph*{Words of bounded size.} A bound on values matters because a machine with words of
that size should then behave the same. \lean{execW w} is \lean{exec} with one change: an
expression with a value of $2^w$ or more, or a subexpression with one, fails to evaluate
(the machine traps on overflow rather than wrapping around), as an
out-of-range array access does (\cref{app:model}). \lean{execW\_agree} says that
$\lean{execW w c s} = \lean{exec c s}$, with the same state, step count and status, whenever
$\lean{dpk c s} < 2^w$. Hence \lean{theoremAB\_word}: for a run of $t$ operations and every
$w \ge 128\lfloor\log_2(t+1)\rfloor + 1286$, the machine with $w$-bit words initializes the
structure, answers every query as the specification does, and takes the same number of
steps as the unbounded machine, so Theorems A and B hold for it verbatim. The theorem fixes $w$ from the
length of the run; the per-operation bound of \cref{fig:main} would also support a machine
whose words widen as the run proceeds, which we do not formalize.

\begin{figure}[t]
\begin{lstlisting}[style=lean]
theorem theoremABW_MW {m₀ : State} {KB : Nat}
    (hkb : m₀.loc "kb" = KB) (h4 : 4 ≤ KB) (h512 : KB < 2 ^ 512)
    (ops : List OM.Spec.Op) (hv : OM.Spec.Valid OM.Spec.State.empty ops) :
    (exec initWS m₀).status = .ok ∧ (exec initWS m₀).cost ≤ initB ∧ -- B (init)
    dpk initWS m₀ < 2 ^ 1029 ∧                                        -- W (init)
    ∃ cs ps, runMW (exec initWS m₀).st ops =
        some (OM.Spec.run OM.Spec.State.empty ops, cs, ps) ∧          -- A
      cs.length = ops.length ∧ ps.length = ops.length ∧
      (∀ c ∈ cs, c ≤ opB KB) ∧                                        -- B
      ∀ i (hi : i < ps.length), ps[i] < 2 ^ (128 * Nat.log2 (i + 1) + 1286) -- W

theorem opB_four : opB 4 = 1133561742
theorem initB_eq : initB = 6648
\end{lstlisting}
\caption{Theorems A, B and W for the machine program (\lean{OM/Impl/PeakRun.lean},
\lean{OM/Impl/TheoremsM.lean}; the comments are ours). \lean{exec c m} runs \lean{c} from
\lean{m} and returns a final state (\lean{.st}), a step count (\lean{.cost}) and a status;
\lean{runMW} runs the operations in turn and returns their answers, their step counts
\lean{cs} and their largest computed values \lean{ps}, each measured by \lean{dpk}.
\lean{initB} and \lean{opB} are the step bounds of initialization and of an operation.
Operations name elements by the specification's handles, the creation indices;
\lean{insertW\_ok} shows that the machine's insertion returns exactly these. The axioms used are \lean{propext}, \lean{Classical.choice} and \lean{Quot.sound}.}
\label{fig:main}
\end{figure}

\subsection{Proof architecture}\label{sec:proof}

\begin{figure}[t]
\centering
\begin{tikzpicture}[
  lvl/.style={draw, rounded corners=2pt, align=center, font=\small, minimum width=3.6cm, minimum height=0.9cm},
  arr/.style={->, thick},
  lab/.style={font=\scriptsize, align=left}]
  \node[lvl] (spec) at (0,5.1) {Specification\\[-1pt]{\scriptsize lists, \lean{Spec.run}}};
  \node[lvl] (f) at (0,3.4) {Level F\\[-1pt]{\scriptsize pure functional implementation}};
  \node[lvl, dashed] (l) at (0,1.7) {Intermediate language\\[-1pt]{\scriptsize transcription of Level F, \lean{evalF}}};
  \node[lvl] (m) at (0,0) {Level M\\[-1pt]{\scriptsize compiled machine program, \lean{exec}}};
  \draw[arr] (f) -- node[right, lab, xshift=4pt] {Theorem A: answers, by the correctness invariant\\cost invariant: no guard fails (\cref{sec:why})} (spec);
  \draw[arr] (l) -- node[right, lab, xshift=4pt] {agreement lemmas, one per Level F function:\\same state, cost at most a stated bound (Theorem B)} (f);
  \draw[arr] (m) -- node[right, lab, xshift=4pt] {\lean{sim}: verified translation,\\same outcome and the same number of steps} (l);
\end{tikzpicture}
\caption{The levels of the proof. Level F and its intermediate-language transcription are written by
hand; Level M is compiled from the transcription. The exception is the wrapper, which is
written directly as machine code and proved against Level F with a weakest-precondition
calculus (\cref{app:proof}).}
\label{fig:levels-proof}
\end{figure}
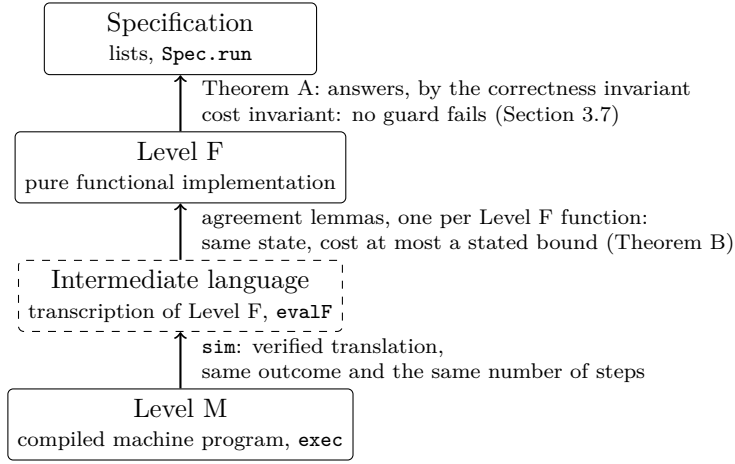

Proving the machine program directly against the specification would mix the mathematics of
the algorithm with memory layout, so the proof goes through three levels below the
specification (\cref{fig:levels-proof}).
\emph{Level F} (functional) is the algorithm written as pure Lean functions on a structured state of
records and arrays, with explicit program counters for the background work. All reasoning
about the algorithm happens there, in two invariants. The \emph{correctness} invariant says
that keys increase along the element list and that its live elements form the list of the
specification; each guarded write keeps it and the fallback restores it, so its proof never
mentions sizes or timing. The \emph{cost} invariant carries the bounds of \cref{sec:why}. It
implies that no guard fails, that the queue's linear-time path is never taken, and that every
list a top job traverses has at most 160 entries. Level F counts no steps. Theorem B combines the agreement lemmas below, which bound each
procedure's steps by a function of the sizes it traverses, with the cost invariant, which
bounds those sizes by constants.

\emph{Level M} (machine) is the machine program, and the cost bound is about it. Relating it to Level F
means relating records and lists to flat arrays of numbers: which array holds which field,
that a write to one array leaves the others unchanged, and that every index stays within its
capacity. Doing this for every procedure would repeat the same memory reasoning each time. Instead, each Level F function is transcribed by hand into a small
\emph{intermediate language}, whose semantics runs directly on Level F states but charges what the
compiled code will cost, and a compiler produces Level M from the transcriptions. The compiler
is a verified change of data representation, from records and lists to flat arrays, that
preserves states and step counts exactly. The intermediate language is tailored to this algorithm:
besides assignments, conditionals, loops and calls, its 24 statement forms include writes to
the structure's own lists and records.
One theorem, \lean{sim}, does all the memory reasoning about a core: for every program, the compiled code
ends exactly as the intermediate-language semantics says, in the same number of steps, unless the
intermediate-language run ends in an internal error (\lean{bug}), which the agreement lemmas exclude. What remains
for each procedure is an \emph{agreement lemma} with its Level F function, proved purely at
Level F. Because \lean{sim} shows that the costs charged in the intermediate language are met by an
actual memory layout on the machine, the interpreter remains the only cost model to trust.
The wrapper, which manages the two cores and the handle maps, is written directly as machine
code and proved against Level F with a weakest-precondition calculus. \Cref{app:formal} gives
the details.

\subsection{An example: the order query}\label{sec:example}
The order query shows every level on a small scale (\cref{fig:order}). At Level F,
\lean{order x y} reads the key of each element, the triple of its chunk's tag, its group's label
and its own label, and compares the two keys lexicographically. Its intermediate-language transcription
\lean{orderP} reads the same fields with field reads (\lean{.eF}, \lean{.gF}, \lean{.cF}) and
computes the comparison as one expression. The agreement lemma \lean{order\_agree} says that
whenever the Level F query returns $b$ and the indices it follows are below the array
capacities, the transcription ends in the same state, with $b$ in its result local \lean{r},
and is charged at most 80 steps. Because both sides are straight-line code, \lean{simp}
unfolds both runs and \lean{omega} closes the remaining comparison of keys. The caller discharges the capacity hypotheses
from the invariants: $x$ and $y$ are listed elements, so their groups and chunks are live units
(by the correctness invariant and the chunk part of the cost invariant), and live units lie
below the capacities. \lean{sim} then turns the intermediate-language run into a run of the compiled machine code
with the same state and step count, and the wrapper's proof, in the weakest-precondition
calculus, adds the steps that translate handles into element indices. Larger procedures follow
the same pattern, but their agreement lemmas need loop invariants, and their capacity
hypotheses come from the cost invariant.

\begin{figure}[t]
\begin{lstlisting}[style=lean]
-- Level F
def key (e : Nat) : CM (Nat × Nat × Nat) := do
  let x ← getE e
  let g ← getG x.grp
  return ((← getC g.chk).tag, g.lab, x.lab)
def order (x y : Nat) : CM Bool := return keyLt (← key x) (← key y)

-- intermediate language
def keyS (e t g l : Reg) : Stm :=
  g ::= .eF .grp e ;; l ::= .eF .lab e ;;
  t ::= .cF .tag (.gF .chk g) ;; g ::= .gF .lab g
def orderP : Proc where
  name := "order"
  params := ["x", "y"]
  rets := ["r"]
  body := keyS "x" "ta" "ga" "ea" ;; keyS "y" "tb" "gb" "eb" ;;
    "r" ::= (("ta" <ₑ "tb") ||ₑ (("ta" =ₑ "tb") &&ₑ
              (("ga" <ₑ "gb") ||ₑ (("ga" =ₑ "gb") &&ₑ ("ea" <ₑ "eb")))))

-- agreement (capacity hypotheses on x, y and their group and chunk elided)
theorem order_agree ... (h : (order x y).run σ = .ok b σ') :
    ∃ L' n, evalF κ orderP.body σ L = .ok σ' L' n ∧ L' "r" = b2n b ∧ n ≤ 80
\end{lstlisting}
\caption{The order query at Level F, in the intermediate language, and the agreement lemma between
them. \lean{CM} is the Level F state and error monad; \lean{Stm} is the type of
intermediate-language statements, \lean{::=} and \lean{;;} are assignment and sequencing, and
\lean{.eF}, \lean{.gF}, \lean{.cF} read element, group and chunk fields. In
\lean{order\_agree}, $\kappa$ gives the array capacities, $\sigma$ and $\sigma'$ are Level F
states, \lean{L} and \lean{L'} the locals before and after (not the bit length $L$), \lean{n}
the steps charged, and \lean{b2n} maps a Boolean to $0$ or $1$ (\lean{OM/Model/Core.lean},
\lean{OM/Impl/Elem.lean}).}
\label{fig:order}
\end{figure}

\subsection{A harder example: room in the top tables}\label{sec:room}
The order query's agreement lemma needs no invariant. The guards of the top tree need one that spans many
operations: a new table entry takes the average of its neighbours' labels, so its guard
succeeds only if that gap has not closed, and a job's traversal of a table is short only if
the table is. Both follow from one bound on \emph{insertion events}. For every table $X$, the
Lean invariant \lean{Room} (\cref{fig:room}) keeps two auxiliary, proof-only quantities: the number $n_0(X)$ of
entries $X$ was created with, and the number $d(X)$ of insertion events it has seen since. An
event adds two entries, so $X$ has at most $n_0(X) + 2d(X)$ entries, and every gap of $X$ is
at least $\lfloor\lfloor 2^\beta/(n_0(X)+1)\rfloor/2^{d(X)}\rfloor$ (\lean{gaps}; the
divisions are integer divisions, as in the fresh labels themselves).

The difficulty is to bound $d(X)$, and here the weights enter. At height $h \ge 2$, an
insertion event in $X$ is a child of $X$ being drained, and the potential \lean{pot2} pays for
it in one of two ways. Write $\gamma = W_{h-1} + 1 - \tau_{h-1}$, where $W_{h-1}$ is the
threshold above which a child is drained and $\tau_{h-1}$ bounds the weight of a node when it
is created. Schematically,
\[ \gamma\, d(X) + \mathit{growth}(X) + \mathit{base} + \gamma\, \mathit{oldLive}(X) \;\le\;
   \mathit{ew}(X) + \gamma\, n_0(X), \]
where $\mathit{growth}(X)$ sums, over the live children created after $X$, their weight above
$\tau_{h-1}$, $\mathit{oldLive}(X)$ counts the live children $X$ was created with, and $\mathit{base}$ is a
lower bound on the initial weight of $X$ itself. When a
child created after $X$ is drained, it had grown by at least $\gamma$, and its contribution to
$\mathit{growth}$ disappears, paying for the new $\gamma$ in $\gamma\, d(X)$. A child that $X$
inherited may already be heavy, so its drain cannot be charged to growth; instead
$\mathit{oldLive}$ falls by one and releases $\gamma$ of the initial credit $\gamma\,
n_0(X)$. The effective weight $\mathit{ew}(X)$ is the weight below $X$, or for a node being
drained the weight below its two new nodes together, and the cost invariant bounds it, so
$d(X)$ stays below a constant $D_{\max}(h)$ (\lean{Room.dr\_lt}). Let $n_{0,\max}(h)$ bound the entries a table of height $h$ is created with (\lean{n0le}).
Two conditions on the parameters, proved for every height when $a = 8$ and $s = 11$, finish
the argument. $2^{D_{\max}(h)}(n_{0,\max}(h)+1) \le 2^\beta$ (\lean{room\_8}) reserves one
halving beyond the events already seen: since $d(X) + 1 \le D_{\max}(h)$ in every state, every
gap is at least two labels wide, so the average taken by the next insertion lies strictly
between its neighbours, and every gap is at least one afterwards. $n_{0,\max}(h) + 2D_{\max}(h) \le 160$ (\lean{tab\_8}) bounds the
table. So the logical order is: weight growth bounds the events, the events bound
the table's length and its halvings, and $\beta$ is chosen to accommodate the halvings.
Preserving \lean{pot2} is where the drain proofs need most care: a drain start creates two
tables at once, a ghost entry goes to the side of its own left new node, and a drained node's
effective weight is that of its two new nodes together. \lean{Room} or its fields are used in
12 of the 46 files of the top-structure proof.

\begin{figure}[t]
\begin{lstlisting}[style=lean]
structure Room (s : TS) (cl : List Nat) (dr n0 : Nat → Nat) : Prop where
  gaps : ∀ X, isN s X → GapsGe s.P.beta (gapB s.P.beta (n0 X) (dr X)) (labs s X)
  n0le : ∀ X, isN s X → n0 X ≤ n0max s.P (ht s X)
  pot1 : ∀ X, isN s X → ht s X = 1 →
    2 * dr X + (if s.P.H < X then lamd s.P 1 else 1) ≤ ew s cl X
  pot2 : ∀ X, isN s X → 2 ≤ ht s X →
    gamd s.P (ht s X - 1) * dr X + growth s cl X
      + (if s.P.H < X then lamd s.P (ht s X) else 1)
      + gamd s.P (ht s X - 1) * oldLive s X
      ≤ ew s cl X + gamd s.P (ht s X - 1) * n0 X

-- in a section with (ha : P.a = 8) (hs : P.s = 11)
theorem room_8 (hb : P.beta = 64) (h : Nat) (h1 : 1 ≤ h) :
    2 ^ Dmax P h * (n0max P h + 1) ≤ 2 ^ P.beta
theorem tab_8 (h : Nat) (h1 : 1 ≤ h) : n0max P h + 2 * Dmax P h ≤ 160
\end{lstlisting}
\caption{Label room in the top tables. \lean{s} is the state of the top structure (the sweep
parameter is \lean{s.P.s}) and \lean{cl} its chunks in order; \lean{isN s X} says that
\lean{X} is a node, live or ghost, \lean{ht} gives its height and \lean{labs} its table
labels, and \lean{GapsGe}~$\beta$~\lean{B} says that every gap is at least \lean{B}. \lean{dr X} counts the insertion
events of table \lean{X}, \lean{n0 X} its initial entries, \lean{gapB} the resulting gap bound
(with integer divisions), \lean{ew} the effective weight below \lean{X} (for a node being
drained, the weight below its two new nodes), \lean{gamd} the growth $\gamma$ that pays for
one event, and \lean{lamd} a lower bound on the weight of a new node. The test
\lean{s.P.H < X} separates nodes created by drains from the initial ones, and \lean{pot1} is the
height-1 form of \lean{pot2}. \lean{room\_8} and \lean{tab\_8} are stated in a section with the
hypotheses \lean{P.a = 8} and \lean{P.s = 11} (\lean{OM/Proof/Top/Inv.lean},
\lean{OM/Proof/Top/TopParams.lean}).}
\label{fig:room}
\end{figure}

\subsection{Memory, value sizes and trust}\label{sec:memwords}\label{sec:trusted}
Two further theorems bound resources. The arrays of a core of capacity $N$ have a total
capacity of at most $40{,}000N$ cells (a bound on core arrays, not on all memory), and a second core being built has capacity at most $4N$; the handle maps, which grow
with the number of insertions, are not included. Theorem W is proved in three steps. The
first bounds every value stored in the cores of Level F, other than handles, by $2^{64L}$,
with the larger $L$ during a rebuild (\lean{theoremW\_stored}); handles and wrapper counters
are bounded by the number of operations. The second bounds the values each intermediate-language
procedure computes, through an extension of \lean{sim} that also relates the largest values,
by $2^{2K+4}$ plus a handle term, where $K = 64L + 128$ covers both the current core and a
core created during the operation. The third extends the bound to the
wrapper's machine code with a weakest-precondition calculus that carries a budget for the
largest value. A new core has $L$ at most $2$ larger, and the capacity is at most $24t + 64$
after $t$ operations; together these give the bound of \cref{fig:main} in $i$ alone. What
remains for a word-RAM result is discussed in \cref{sec:limits}.

The top-level theorems use only Lean's standard axioms \lean{propext}, \lean{Classical.choice}
and \lean{Quot.sound}. To assess what they mean, a reader must check the specification,
the machine interpreter (\lean{OM/Machine/Syntax.lean} and \lean{Exec.lean}), the statements
of \cref{fig:main}, and the definitions that
connect them: \lean{runMW} runs the operations in turn, \lean{opS} encodes an operation and
its handle arguments as machine code, \lean{ansOf} reads an answer from the machine state, and
\lean{dpk} reads the largest value the interpreter recorded. The statement on $w$-bit words
adds \lean{execW}, which repeats \lean{exec} with a bounded evaluator, and \lean{runMw}, which runs the
operations as \lean{runMW} does, with \lean{execW} in place of \lean{exec}, and returns their
answers and step counts; its statement also equates the $w$-bit run with \lean{runM}, the run of
\cref{fig:main} without the peaks. The model
assigns unit cost to arithmetic, shifts and $\lfloor\log_2\rfloor$ (\lean{msb}) on unbounded
words and to zero-filled allocation
\cite{BriggsTorczon93}; Theorem W bounds the size of those words. The Rust transpiler and runtime are outside the proof. The script \texttt{scripts/reproduce.sh}
in the repository repeats the build, the axiom audit of the main theorems, a fresh
transpilation and the cross-check of \cref{app:eval}.
\Cref{app:formal} gives the machine semantics and the size of the development.

\subsection{Proof engineering}\label{sec:engineering}
The development uses Lean~4.35 and Mathlib, for real numbers (the harmonic numbers of the
timing inequalities), list lemmas and arithmetic tactics such as \lean{ring} and
\lean{norm\_num}. The proofs are mostly \lean{simp}, \lean{omega} and
\lean{decide} (about 11{,}000, 4{,}700 and 1{,}500 uses), and \lean{native\_decide} is never
used. A clean build of the development, with Mathlib prebuilt, takes about 8 minutes on 24
cores; the 203 modules the library imports take 13 minutes of CPU time in total, so the build is limited by its longest chain of
imports rather than by the cores. Four lessons stand out.

\emph{A contract separates the invariants.} The invariants of \cref{sec:why} depend on one
another and are proved together. In Lean the argument is ordinary induction over operations, on one invariant,
\lean{CostInv} (\cref{fig:contract}), which combines the invariant of groups, chunks and the queue, the state of the zeroing game
with its accounting of rounds, and an abstract invariant \lean{TI} of the top structure. The
proofs about groups, chunks and the queue are made against a contract, \lean{TopContract TI}~$\mu$~\lean{J cap}: starting a chunk
split yields a job whose remaining work $\mu$ is at most $J$, and every job step keeps
\lean{TI} and decreases $\mu$. The top structure is proved separately to satisfy the contract,
with $\mu$ the number of steps left in its running job. Neither proof looks inside the other,
and the two were developed in sequence.

\begin{figure}[t]
\begin{lstlisting}[style=lean]
structure CostInv (TI : Core → Prop) (μ : Core → Nat) (J n R : Nat)
    (L : List Nat) (s : Core) : Prop where
  low  : LowInv s L    -- groups, chunks, queue; s holds the sequence L
  ti   : TI s          -- the top structure's invariant, abstract
  game : ∃ y i a ip js, GameInv n R s y i a ip ∧ RoundOK s.P.K_R J μ s ip js

structure TopContract (TI : Core → Prop) (μ : Core → Nat) (J : Nat)
    (cap : Params → Nat) : Prop where
  frame  : ∀ s s', TopSame s s' → TI s → TI s'
  mframe : ∀ s s', TopSame s s' → μ s' = μ s
  start  : ∀ s A x, INV s A → TI s → s.resJob = none → LiveC s x →
    cSt s x = .quiet → s.C.size + 2 ≤ cap s.P →
    ∃ job s2, (insertAround x ...).run ... = .ok job s2 ∧
      1 ≤ μ (markState s2 ...) ∧ μ (markState s2 ...) ≤ J ∧
      TI (markState s2 ...)
  step   : ∀ s A job, INV s A → TI s → s.resJob = some job →
    ∃ r s', (jobStep job).run s = .ok r s' ∧ TI { s' with resJob := r } ∧
      ∀ j', r = some j' →
        1 ≤ μ { s' with resJob := r } ∧ μ { s' with resJob := r } < μ s
\end{lstlisting}
\caption{The cost invariant and the contract of the top structure (arguments elided, comments
ours). In \lean{CostInv}, \lean{L} is the specification's sequence (not the bit length $L$) and
\lean{s} the core; \lean{n} and \lean{R} are the number of counters and the insertions per
round of the zeroing game, \lean{GameInv} is the game's invariant for the current round
(round-start values \lean{y}, the counter \lean{i} just reset, and the additions \lean{a} of
the \lean{ip} insertions since), and \lean{RoundOK} accounts for the \lean{js} job steps of
the round against $J$. In \lean{TopContract}, \lean{cap} bounds the number of chunks the tree
has room for, \lean{resJob} is the running job, and \lean{markState} is the state after the
chunk split starts. \lean{frame} and \lean{mframe}: steps
that leave the top structure's state unchanged (\lean{TopSame}) keep \lean{TI} and $\mu$.
\lean{start}: entering two chunks around a quiet chunk \lean{x} starts a job with at most
$J$ steps of work. \lean{step}: a job step keeps \lean{TI} and, unless the job ends, decreases
$\mu$. \lean{INV} is the correctness invariant (\lean{OM/Proof/Cost/CoreCost.lean},
\lean{OM/Proof/Cost/TopIface.lean}).}
\label{fig:contract}
\end{figure}

\emph{State-based measures cost constants.} The pen-and-paper proof bounds label room through
the history of each gap. Lean replaces histories by measures of the current state: one halving
counter per chunk instead of halving histories per gap, and a potential over table weights
instead of the histories of table insertions. Each invariant then becomes a property of one
state, preserved by each step, at the price of coarser bounds: group labels grew from about
$9L$ to about $30L$ bits and $\beta$ from 48 to 64. $K_R$ grew from 4 to 8 for a different
reason: with 8, the waiting-chunks inequality follows for every $L \ge 8$ from the bound
$\mathcal{H}_m \le 1 + \ln m$; with 4 it holds, but only numerically, with about 5\% to
spare at $L = 8$.

\emph{Prove memory reasoning and peak tracking once.} \lean{sim} is the only theorem about
the layout of a core's arrays, and the peak bound reuses its structure: it was extended once
(\lean{sim\_peak}) to relate the
largest values of the two runs, after which the bounds on computed values are proved, like the
agreement lemmas, at the level of the intermediate language. Only the wrapper is reasoned about as
machine code.

\emph{Where the lines go.} The top structure takes 14{,}807 lines, about half of them for
showing that starting and advancing a drain preserves the tree's shape, order and timing
invariants: a drain start changes several nodes, tables and paths at once. The bound on
computed values takes about 12{,}000 lines, most of them a second pass over every procedure that
tracks the width of each expression. The guards pay off in the size of the correctness proof:
8{,}421 lines, against about 29{,}000 for the costs at Level F, because it never mentions timing.

\section{Execution and Practical Cost}\label{sec:eval}

Two Rust artefacts serve different purposes. The first is transpiled statement for statement
from the verified machine program by an unverified transpiler; it uses arbitrary-precision
values, counts machine steps, and agrees with Lean's machine semantics on answers, step counts
and the bit lengths of the largest computed values on every cross-checked run (\cref{app:eval}). The second, a native
implementation, implements the same algorithm with records and fixed-width labels. It is not
verified, but it and Level F both reproduce 28 traces written by our Python prototype, 14.6
million operations including deletion-heavy workloads and repeated rebuilding. This evidence
checks the tested executions; it does not prove the transpiler, runtime or native
implementation correct.

\Cref{tab:eval-summary} gives representative results. On insertions the native
implementation's mean time is 1.3--2.8 times the AVL tree's. At the 99.99th percentile of a
run it is as fast as the tree or faster (0.8--1.0 times), but the tree has the lower slowest
repeatable insertion time: by that statistic, native insertions take about 1.3--2.3 times as
long, a ratio that varies between sessions and seeds. Native cold order queries are 2.2--6.3 times faster than the tree's, by either estimator,
and native needs about 8.6 times the memory. Its advantage over the amortized baseline is in
the worst case: clustered insertions force that baseline into relabelings of about 5\,ms,
while on average the amortized baseline is faster than both. The transpiled program, which computes with arbitrary-precision words and counts every
step, is much slower than all three (\cref{app:eval}). These are practical trade-offs, not a
claim that the asymptotically stronger bound wins at every size or workload.

\begin{table}[t]
\centering
\small
\caption{Selected measurements in $\mu$s: mean / 99.99th percentile of a run / slowest
repeatable time (the largest per-operation minimum over three runs; see the text). Insertions are
those that grow the sequence from $10^7$ to $10^8$ elements; cold queries are 500 random pairs
in the same range after random insertions, each measured in two cold runs after a warm run
that selects the measured operations (too few for a percentile). For cold queries, the mean is the
mean over queries of each query's minimum over the two runs, preferring a run without a page
fault and leaving out queries that fault in both, and the slowest time is the largest of these
minima; the means of the individual runs are 1.02--1.03 (native), 6.4--6.5 (AVL) and
0.56--0.57\,$\mu$s (amortized). Memory is peak resident bytes per element after $10^8$ random insertions.
Hardware: Intel Core Ultra 9 285K, Linux 7.0, one performance core.}
\label{tab:eval-summary}
\begin{tabular}{lrrr}
\toprule
 & native (unverified) & AVL & amortized \\
\midrule
Random insertion & 1.00 / 5.3 / 12.1 & 0.77 / 5.5 / 7.2 & 0.54 / 12.3 / 15.3 \\
Clustered insertion & 0.085 / 2.9 / 6.3 & 0.030 / 3.2 / 3.6 & 0.022 / 3.4 / 5{,}050 \\
Cold order query & 0.93 / -- / 1.4 & 5.9 / -- / 8.4 & 0.50 / -- / 0.88 \\
Memory (bytes/element) & 112 & 13 & 21 \\
\bottomrule
\end{tabular}
\end{table}

Each deterministic workload runs three times. The slowest repeatable time takes each
operation's minimum over the runs. This removes pauses that hit an operation in only some runs,
but it is not a measured worst-case latency: a recurring cost that falls on different operations
in different runs is removed too. The maximum of a single run is a different quantity, and
it varies from run to run. In the session of the table, it is about 0.3\,ms at $10^7$ elements
for every implementation in most runs, with occasional runs at 2.7--4.3\,ms. At $10^8$ it is
4.3--20.5\,ms for native random insertion, 4.3--6.1\,ms for the tree and 4.3--7.0\,ms for the
amortized baseline, and one of three native clustered-insertion runs has a 92\,ms outlier that
the other two do not show. The same values (0.3, 2.7, 4.3 and 6.1\,ms) recur across
implementations and workloads, even in order queries. A test that only writes to fresh memory
pages shows one source: about one first touch in 2{,}000 stalls for about 0.3\,ms. Native
touches about 8.6 times as much memory as the tree and meets more of these stalls: at
$10^8$, 733--900 random insertions per run take over 100\,$\mu$s, against 28--179 for the tree
and 111--178 for the amortized baseline. The amortized
baseline's relabeling under clustered insertion is the one maximum that recurs at the same
operation in every run, which identifies it as algorithmic; we do not claim that every other
outlier has a memory-system cause. None of the implementations
has a proved wall-clock bound.

With $10^3$ elements and $10^9$ operations, alternating random deletions and insertions, the
times per operation stay the same over the whole history, and memory grows by 12 bytes per
insertion for native, 13 for the AVL tree and 20 for the amortized baseline: in each case it
is the map from handles, which are never reused. Native's cores in use stay below 1\,MB through
$2.7 \cdot 10^5$ rebuilds.

The transpiled program's largest measured step count grows with the size even though its
proved bound is constant: from $1.3 \cdot 10^5$ steps at $10^5$ elements to $3.1 \cdot 10^5$ at
$5 \cdot 10^7$, against the proved $1.1 \cdot 10^9$. The number of top-job steps per operation
is fixed, but as the tree gets taller more of them start a drain, the most expensive kind of
step. That number is bounded by the job steps per operation, so the growth must stop. The workload
definitions, deletion results and the explanation of this growth are in \cref{app:eval}.

\section{Limitations}\label{sec:limits}

\paragraph*{From bounded values to a word RAM.}
The machine's numbers are unbounded, and unit-cost arithmetic on unbounded words makes
random-access machines unrealistically powerful~\cite{HartmanisSimon74}. Theorem W rules this
out: after $t$ operations no value has more than $128\lfloor\log_2(t+1)\rfloor + 1286$ bits, and
a machine with words of that size runs the program with the same results and step counts
(\lean{theoremAB\_word}). That machine still has unit-cost multiplication, division, shifts
and $\lfloor\log_2\rfloor$, which common word-RAM variants provide, and zero-filled allocation
of arrays of any size, which they do not (see \emph{Allocation} below). On a word RAM with
words of $\Theta(\log t)$ bits, a value of that size spans a constant number of words (about
128), so each unit-cost operation on it takes a constant number of word operations, with
schoolbook multiplication and division; only the constant grows. The constant factor in the
word size is large, because the proof bounds every stored value by the same $64L$ bits, well
above the label widths of \cref{app:params}, and products double that; the native
implementation needs far fewer bits (\cref{app:eval}). Only handles and a few wrapper counters
grow with the number of operations rather than with the capacity, handles because they are
never reused; like any counter that grows by one per operation,
they fit in 64 bits for every feasible run~\cite{ClochardEtAl15}.

\paragraph*{Space and the length of the sequence.}
The bounds are in terms of the number of operations and the capacity, not the number $n$ of
live elements. Rebuilding shrinks a core whose live elements fall below an eighth of its
capacity, so in practice the capacity follows $n$, but we prove only that it is at most
$24t + 64$ after $t$ operations, and no bound in terms of $n$. The handle maps grow with the
number of insertions and the cores with the capacity, so the space grows with the number of
operations rather than with $n$; the memory theorem covers the cores only. Words of $O(\log n)$ bits and $O(n)$ space would need
handles to be reused: a free list of handles, with a per-handle record that a rebuild
updates, and a proof that the capacity tracks $n$. We have not attempted either.

\paragraph*{Allocation.}
The machine allocates a zero-filled array of any size in one step, and the program relies on
it: a new core allocates arrays for up to $40{,}000N$ cells, far more than it uses, and every
rebuild allocates handle maps with room for every handle issued so far, about $t$ cells after
$t$ operations, whatever the length of the sequence. A standard
word RAM has no such instruction. The usual simulation initializes memory lazily, with
constant overhead per access and a constant factor in
space~\cite{AhoEtAl74,BriggsTorczon93,HagerupKammer17}, but we do not prove it, and the Rust
runtime instead relies on lazily mapped operating-system memory. Since every rebuild allocates
fresh arrays, the addresses such a simulation touches would grow with the total size allocated
over the run unless freed arrays were reused.

\paragraph*{From machine steps to execution time.}
The model gives zero-filled allocation unit cost and does not account for caches, page
faults, reclamation or scheduling. The Rust runtime approximates allocation with lazily
mapped memory and incremental cleanup; it is unverified. The native implementation and the
Rust transpiler are also unverified. The measured native memory includes retained retired
cores, which the formal core-memory bound does not cover. Finally, the theorems cover valid
histories only: they do not specify behaviour for deleted or foreign handles.

\section{Related Work}\label{sec:related}

Dietz and Sleator's worst-case construction supplies the main techniques: incremental
splitting, largest-first service and the zeroing game
\cite{DietzSleator87,DietzSleator88}. Bender et al.\ give a simpler amortized algorithm and
connect labeling to weight-balanced trees~\cite{BenderEtAl02}, and Kopelowitz replaces
Willard's algorithm at the top by a weight-balanced B-tree~\cite{Kopelowitz12}. We keep this
approach to grouping, scheduling and the top tree, but guard every write and make the tree's
labels immutable, with ghost drains instead of relabeling in place. As far as we know, no
worst-case $O(1)$ order-maintenance algorithm has been implemented before, and no
order-maintenance algorithm has been verified in a proof assistant. \Cref{sec:bg-prior} gives the historical comparison.
Guo and Sekerinski face the same need for queries to stay correct while labels are being
changed, in a concurrent amortized structure~\cite{GuoSekerinski26}: they relabel in place,
installing precomputed labels in an order that keeps every intermediate state sorted, so that
order queries can run lock-free. Our construction avoids relabeling in place altogether.

The closest verified relatives are real-time deques: T\'oth and Nipkow verify Chuang and
Goldberg's deque in Isabelle~\cite{TothNipkow23}, and Viennot et al.\ verify Kaplan and
Tarjan's catenable deques in Rocq~\cite{ViennotEtAl25}. Tóth and Nipkow argue constant time
informally, from the non-recursive operations of their verified deque. Viennot et al.\ verify
functional correctness and separately provide a non-recursive OCaml implementation; their
verified code keeps computations on indices, so neither it nor its extraction attains the
constant bound, and the two are not formally connected. Madhavan et al.\ do verify
concrete worst-case step bounds for Okasaki's real-time queue and deque, at most 40 and 893
steps, by SMT-based contract checking in Leon against an instrumented source semantics whose
soundness is proved on paper~\cite{MadhavanEtAl17}. As there, our bound does not follow from
the syntax, but here the invariants it rests on concern the whole structure: the loops of a top
job are bounded only by invariants on table sizes, which hold because of the timing arguments
of \cref{sec:why} about weights, waiting chunks and drains. We prove it in Lean, from the step counts of a deeply
embedded machine. Other frameworks support mechanized cost reasoning through time credits,
cost monads or refinement to lower-level code
\cite{GueneauEtAl18,Danielsson08,HaslbeckLammich21,NiuEtAl22}. Like Haslbeck and
Nipkow~\cite{HaslbeckNipkow18}, we count steps in the semantics of a deeply embedded language;
we do so for a complete data structure, with a verified translation from records to arrays
that preserves step counts.
\Cref{app:related} compares these proof approaches in more detail.

\section{Conclusion}\label{sec:concl}

Worst-case constant-time order maintenance has been known for almost four decades, from
descriptions that are intricate and, in places, incomplete. We assembled an algorithm from the
published components, adapted it for proof by guarding every write and by never changing the
labels of existing entries in the top tree's tables, and proved in Lean that a program for it, run on a random-access machine
whose interpreter counts every step, answers queries correctly and takes at most a fixed number
of steps per operation, and that it runs unchanged on a machine whose words have $O(\log t)$
bits after $t$ operations. Checking the
literature exposed errors in two of the papers we build on or compare with, and a pen-and-paper proof written
before the formalization found a bug and two gaps in our own design that extensive testing had
missed. The transpiled program is slow, but the algorithm is not: implemented natively, it
stays within a small factor of an AVL tree.

What remains for the textbook word-RAM statement is to replace unit-cost allocation of large
arrays, to reuse handles and prove that the capacity tracks the number of live elements, so
that word size and space depend on the length of the sequence rather than on the number of
operations, and to shrink the constants (\cref{sec:limits}). Beyond that, the native
implementation could be verified, by refining it to the
machine program or by targeting a lower-level language, and the proved constant could be
tightened. More broadly, guarded writes may help with other deamortized structures: they make
the correctness invariant a local property of each write, and leave to timing arguments only
the cost and the proof that the fallback never runs.

\section*{AI Disclosure}

This work was done in collaboration with the large language models Claude Opus~5.5
(Anthropic) and Astra~6 (OpenAI), used as coding and writing assistants. They contributed to the
design of the algorithm, the pen-and-paper proof, the Lean development, the Python prototype,
the Rust implementations, the experiments and the text of this paper. Lean checks the proofs.
The author directed the work and checked by hand the trusted parts (the specification, the
machine interpreter, the statements of the main theorems and the definitions they rely on;
\cref{sec:trusted}), the claims about earlier work against the original papers, the
experimental results and this text.

\bibliography{refs}
\clearpage
\appendix

\section{Parameter Choices and Timing Inequalities}\label{app:budgets}

This appendix lists the seven inequalities between the parameters of \cref{tab:params} that
the timing proof of \cref{sec:why} needs, in simplified form; the exact forms, with all
rounding, are proved in Lean for every $L \ge 8$.
\begin{enumerate}
\item \emph{A group split finishes before its siblings fill up} (\cref{sec:split,sec:why}):
  $\lfloor (G + k_G + 3)/2 \rfloor < G$ with $k_G = \lfloor G/3\rfloor + 1$.
\item \emph{Labels never run out} (\cref{sec:keys}). If, in the lifetime of a group or chunk,
  a gap between two labels can be halved at most $U$ times and at most $P$ members are pushed
  at one end, its labels need $B = U + \lfloor\log_2 P\rfloor + 2$ bits (\cref{app:params}). The previous item bounds $U$
  and $P$ for groups by $O(L)$; the next two do so for chunks.
\item \emph{A chunk that has just split may wait} (\cref{sec:reserve}). When a chunk's split
  finishes, each half holds at most $c_{\max} \approx (T + k_C)/2 \approx 1.5L^2$ elements,
  where $k_C$ is the number of insertions during a chunk split. This must not exceed the queue threshold:
  \[ c_{\max} \le \theta = 2L^2, \]
  so a chunk enters the queue only by growing past $\theta$, and the zeroing game starts with
  its counter at $0$.
\item \emph{A waiting chunk is served before it reaches $T$} (\cref{sec:reserve,sec:why}):
  $\theta + R \cdot (\mathcal{H}_{m-1} + 1) \le T$ with $R = \lfloor (13H + 2)/K_R\rfloor + 1$.
\item \emph{A drain finishes before its new nodes are too heavy} (\cref{sec:top,sec:why}):
  $\bigl(\tfrac12 + \tfrac{s}{a(s-2)} + \tfrac{2}{s-2}\bigr) \cdot W_h + O(1) \le W_h$ with
  $W_h = 2a^h$.
\item \emph{Tables and the tree have room} (\cref{sec:top}). A table receives a bounded number
  of entries between two neighbours before its node drains, so a fixed label width suffices:
  the proof needs $\beta \ge 62$, and we take $\beta = 64$. The height must leave room for
  every chunk the core can hold: $2a^{H-1}$ exceeds the number of chunks, so not even a node of
  height $H-1$ ever drains.
\item \emph{A rebuild finishes before the old core fills} (\cref{sec:wrapper}). A rebuild
  starts by the time the core holds $N/2$ elements and copies at $K_B = 4$ steps per update, so
  it ends while the core holds at most about $2N/3 < N$. The new core's capacity is chosen so
  that it holds at most a quarter of it when it takes over, and is at most $4N$.
\end{enumerate}

\section{Label Widths}\label{app:params}

A label range of $B$ bits receives labels by the three primitives of \cref{sec:guards}: the
centre $2^{B-1}$, a push of $\pm D$ next to an end member, and the average of two neighbours.
Two counts determine $B$ and $D$: the most \emph{halvings} a gap can suffer, $U$, and the most
\emph{pushes} at one end, $P$. If $D \ge 2^U$, a gap of size $D$ remains nonzero after at most $U$ halvings,
and if $P \cdot D < 2^{B-1}$, pushes never leave the range. \lean{proofParams} takes the
smallest such values, $D = 2^U$ and $B = U + \lfloor\log_2 P\rfloor + 2$. The proof keeps
auxiliary counters of halvings and pushes per group and per chunk, and shows that they never exceed
the bounds below.

\paragraph*{Element labels.} With $k_G = \lfloor G/3\rfloor + 1$ (the insertions a split lasts)
and $g_{\min} = \lfloor (G-2)/2 \rfloor$ (the smallest size of a group after its split), a
group suffers at most $U_G = G + k_G + \max(0, k_G - g_{\min})$ halvings in its life and at most
$G + k_G$ pushes at one end, so $B_e = U_G + \lfloor\log_2(G + k_G)\rfloor + 2$.

\paragraph*{Group labels.} A chunk receives at most $I_C = T + 2k_C$ insertions in its life,
where $k_C$ bounds the insertions a chunk split lasts. Each split start of a group in the chunk
halves at most two gaps. Split starts are of two kinds: \emph{child} splits, of groups born in
the chunk, at most $\lfloor I_C/g_{\min}\rfloor$ because each is charged $g_{\min}$ insertions;
and \emph{root} splits, of groups that arrived with a chunk split or were the chunk's first,
at most $3M_u + 1$, where $M_u$ bounds the moves of a chunk split (each move takes one group,
or a whole splitting family of at most three). So $U_C = 2(\lfloor I_C/g_{\min}\rfloor + 3M_u + 1)$, pushes at one end are
at most $P_C = 3M_u + (3M_u + 1) + \lfloor I_C/g_{\min}\rfloor$, and $B_g = U_C +
\lfloor\log_2 P_C\rfloor + 2$. The pen-and-paper proof follows each gap's own halving history
and needs only about $9L$ bits; the Lean proof counts all halvings of a chunk together, which is
simpler and still $\Theta(L)$.

\paragraph*{Concrete widths.} Chunk tags take $H\beta = 64(\lfloor L/3\rfloor + 1)$ bits.
\begin{center}
\small
\begin{tabular}{rrrrr}
\toprule
$N$ & $L$ & $B_e$ & $B_g$ & $H\beta$ \\
\midrule
$< 2^8$ & 8 & 28 & 300 & 192 \\
$10^7$ & 24 & 73 & 779 & 576 \\
$10^8$ & 27 & 81 & 869 & 640 \\
$10^{10}$ & 34 & 99 & 1{,}079 & 768 \\
\bottomrule
\end{tabular}
\end{center}
Asymptotically $B_e \approx 2.7L$, $B_g \approx 30L$ and $H\beta \approx 21L$.

\section{Machine Semantics and Proof Architecture}\label{app:formal}
This appendix gives the details omitted from \cref{sec:formal}: the machine language and its
interpreter, the simulation theorem, the program logic of the wrapper, where the numerical
bound comes from, and the size of the development.

\subsection{The machine}\label{app:model}
\Cref{fig:syntax} shows the language. Words are natural numbers. Memory is a finite set of
named arrays, each with a capacity; an access beyond the capacity is a machine error. Each
procedure call has its own frame of local variables. A loop \lean{for x < e do s} evaluates
its bound once, before the first iteration. There is no \lean{while}, and there is no
recursion: a \lean{call} carries the callee's body as a subterm, so a procedure that calls
itself would have an infinite syntax tree. Every program therefore terminates, and the
interpreter
\begin{center}
\lean{exec : Stmt → State → Out}, \qquad \lean{Out} $=$ final state $\times$ cost $\times$ status
\end{center}
is a total, structurally recursive Lean function (\cref{fig:exec}): no fuel, no step-indexed
relation. The status
is \lean{ok}, \lean{fail} (a failed guard, caught by the nearest \lean{catch}, which runs
its handler) or \lean{err} (a machine error, which the proofs exclude). In the machine program
the fallback is omitted and a failed guard aborts the operation; the cost invariant shows
that no guard fails.

The machine with $w$-bit words, \lean{execW w} (\lean{OM/Machine/ExecW.lean}, 77 lines),
repeats \lean{exec} case by case. The only change is that it evaluates every expression with
\[ \lean{e.evalW w s} \;=\; \textbf{if}\ \lean{e.maxv s} < 2^w\ \textbf{then}\ \lean{e.eval s}\ \textbf{else}\ \lean{none}, \]
and argument lists likewise, so a value that does not fit in a word is a machine error.
Every value the program handles is the value of an evaluated expression (locals and memory
cells are read only through expressions, and a loop counter stays below the loop's bound), so
\lean{execW w} never holds a value of $2^w$ or more that it computed.

\begin{figure}[t]
\begin{lstlisting}[style=lean]
inductive Expr where
  | lit (n : Nat) | var (x : Reg) | load (a : Arr) (i : Expr)
  | bin (op : BinOp) (e₁ e₂ : Expr) | un (op : UnOp) (e : Expr)
    -- BinOp: + - * / % <<< >>> < ≤ = ∧ ∨ ;  UnOp: ¬ msb

inductive Stmt where
  | skip | assign (x : Reg) (e : Expr) | store (a : Arr) (i e : Expr)
  | alloc (a : Arr) (n : Expr)              -- capacity n, all cells 0, one step
  | seq (s t : Stmt) | ite (c : Expr) (s t : Stmt)
  | loop (x : Reg) (n : Expr) (body : Stmt) -- for x < n do body
  | call (name : String) (params : List Reg) (body : Stmt) (rets : List Reg)
      (args : List Expr) (dsts : List Reg)  -- the callee's body is a subterm
  | fail | catch (s h : Stmt)
\end{lstlisting}
\caption{The machine language (\lean{OM/Machine/Syntax.lean}, abridged).}
\label{fig:syntax}
\end{figure}

\begin{figure}[t]
\begin{lstlisting}[style=lean]
def exec : Stmt → State → Out
  | .assign x e, s =>
    match e.eval s with
    | some v => ⟨(s.bump (e.maxv s)).setLoc x v, 1 + e.cost, .ok⟩
    | none   => ⟨s, 1 + e.cost, .err⟩
  | .loop x n b, s =>
    match n.eval s with
    | some k => Out.bind ⟨s.bump (n.maxv s), 1 + n.cost, .ok⟩
                  (runFor (exec b) x 0 k)
    | none   => ⟨s, 1 + n.cost, .err⟩
  ...
\end{lstlisting}
\caption{Two cases of the interpreter. \lean{e.cost} is the number of nodes of \lean{e};
\lean{runFor} charges one step per iteration plus the body. \lean{bump} records the largest
value computed (\cref{sec:main}).}
\label{fig:exec}
\end{figure}

\subsection{From Level F to the machine}\label{app:proof}
The intermediate language has assignments, writes to record fields, appends, sequencing,
conditionals, bounded loops, calls, \lean{fail} and \lean{catch}. Its semantics \lean{evalF}
runs directly on Level F states and charges for each statement what the compiled machine code
will cost. An encoding relation \lean{Enc p κ s m} states that machine state \lean{m} holds
core \lean{s} in slot \lean{p}, one array per record field, with capacities \lean{κ}. The
simulation theorem covers every program:
\begin{lstlisting}[style=lean]
theorem sim (c : Stm) : ∀ {s : Core} {L : Loc} {m : State}, Enc p κ s m → m.loc = L →
    Agree p κ (evalF κ c s L) (exec (c.compile p κ) m)
\end{lstlisting}
where \lean{Agree} says that both end with the same status, the machine state holds the new
core, the locals agree, and \emph{the step counts are equal}, except when \lean{evalF} ends in
\lean{bug}, for which \lean{Agree} holds trivially. In words: whatever an intermediate-language
program does under \lean{evalF}, unless it ends in \lean{bug}, its compiled code does on the
machine, in the same number of steps; the agreement lemmas show that the procedures never end
in \lean{bug}.

What remains for each procedure involves no memory. Each Level F function is transcribed
into the intermediate language, and an \emph{agreement lemma} shows, by reasoning purely at Level F,
that on every successful run of the function the transcription reaches the same state within
a stated number of steps. Combined with \lean{sim}, this gives the compiled procedure's result
and its step count. A write the encoding cannot represent makes \lean{evalF} return an error,
so the agreement lemmas also prove that every index stays within its capacity; the capacities
themselves are bounded by functions of $N$ that are further invariants (the number of
top-tree nodes, for instance, is bounded by a potential argument).

\paragraph*{The wrapper and the program logic.}
The wrapper is written directly as machine code. An encoding \lean{WEnc om m p} places the
current core in slot \lean{p}, the next core in slot \lean{1 - p}, a few scalar cells, and
the handle maps; the compiled core code of a slot leaves the other slot untouched. Each
wrapper operation is proved against its Level F counterpart, under the Level M
invariant \lean{MInv om Sp}: Level F's invariant relating the wrapper state to the
specification state \lean{Sp}, together with the capacity conditions of both cores. These
proofs use a weakest-precondition calculus on the machine. Because \lean{exec} is total and deterministic, a weakest precondition is simply a predicate
on its result:
\begin{lstlisting}[style=lean]
def wp (c : Stmt) (Q : State → Nat → Prop) (s : State) : Prop := (exec c s).Ok Q
\end{lstlisting}
where \lean{Out.Ok Q} holds if the status is \lean{ok} and \lean{Q} holds of the final state
and the cost. Each construct has a rule; most are equivalences, those for loops and
\lean{catch} are implications;
for example, \lean{wp (seq c d) Q s} unfolds to \lean{wp c (fun s' n => wp d (fun s'' m => Q
s'' (n + m)) s') s}. The loop rule takes an invariant indexed by the iteration and a cost bound
$c$ for one iteration, and concludes a bound of $1 + \mathit{cost}(e) + k(1 + c)$ for a loop
whose bound expression $e$ evaluates to $k$. Arrays are named globals, so there is no aliasing between different arrays and
no need for separation logic; aliasing within an array is a case split on indices.

\subsection{The numerical bound}
The constant is large, and it is worth saying where it comes from. One insertion into a core
costs at most $94{,}462{,}311$ steps. Almost all of it is the $K_R = 8$ steps of the top
structure, each bounded by $11{,}806{,}996$ machine steps, the sum of the bounds of a job step
($11{,}630{,}300$, of which $8{,}977{,}738$ allow for starting a drain, whose loops nest over
tables and child lists of up to 160 entries each) and of the start of a chunk split
($176{,}350$). An operation performs its own insertion, possibly a
mirrored one, and up to $K_B + 1$ rebuild iterations (the iteration that sets the copy cursor
spends no budget), each of which may copy an element by an insertion into the next core; the
bound adds the insertion and deletion cases rather than taking their maximum. The measured
worst case is about $3 \cdot 10^5$ steps (\cref{app:eval}). We made no attempt to tighten the constant: the theorem is
about the existence of a bound independent of $N$, and the numeral makes that bound concrete.

\subsection{Size of the development}\label{sec:size}
\Cref{tab:size} breaks down the development. Most of it concerns cost: the cost proofs at
Level F and the descent to the machine take about 50,000 lines, the correctness proof about
8,400.

\begin{table}[t]
\centering
\caption{Size of the Lean development (lines, including comments). The counts document the
scale of the development, not the best possible decomposition.}
\label{tab:size}
\begin{tabular}{lr}
\toprule
Part & Lines \\
\midrule
Specification & 124 \\
Machine model and program logic & 896 \\
Level F (functional implementation) & 1,545 \\
Correctness (Level F part of Theorem A) & 8,421 \\
Cost, lower levels and wrapper (Level F) & 14,503 \\
Cost, top structure (Level F) & 14,807 \\
Intermediate language, compiler, machine program and its proofs (Level M) & 20,787 \\
Stored-value bound (Theorem W, Level F) & 4,227 \\
Computed-value bound and $w$-bit words (Theorem W, Level M) & 12,293 \\
Rust transpiler and cross-check & 232 \\
Tests & 134 \\
\midrule
Total & 77,969 \\
\bottomrule
\end{tabular}
\end{table}

\section{Errors and Gaps Found}\label{sec:errors}

\paragraph*{In the literature.} We checked every claim below against the originals: the
technical report version of Dietz and Sleator~\cite{DietzSleator88}, which has the appendix
with pseudocode, the ESA version of Bender et al.~\cite{BenderEtAl02} and the arXiv version of
Kopelowitz~\cite{Kopelowitz12}.
\begin{itemize}
\item \emph{Dietz and Sleator, appendix.} \lean{PlaceAfter(x,y)} gives $y$ a slot after its
  sibling $x$; it tests whether $x$ is the \emph{first} child of its parent where it must test
  for the last, so with $x$ last it reads the slot of a sibling that does not exist, and with
  $x$ first it can compute a slot beyond the next sibling's (their Fig.~3, p.~17). The test reads
  ``\texttt{if x = first(children(p)) then index(y) $\leftarrow$ index(x) + 1}''.
  \lean{PlaceBefore} has the symmetric error. \lean{IncrementalSplit(x)} moves the first child
  of $x$ after the last child of $\mathit{pred}(x)$ (their Fig.~5, p.~18), but on the first move
  $\mathit{pred}(x)$ is the empty node just created by \lean{SetUpSplit} (their Fig.~4) or, at
  depth one, by the reservation process (their Fig.~6), which has no last child. Each error is
  easy to repair once seen; we mention them because the appendix is, as far as we know, the
  only published pseudocode of a worst-case $O(1)$ algorithm.
\item \emph{Bender et al., Lemma~3.} Write $\tau$ for their density parameter (their $T$).
  After a range of $2^i$ labels is relabeled, each half
  holds at most $2^{i-1}\tau^{-i}$ elements and overflows above $2^{i-1}\tau^{-(i-1)}$, so the
  number of insertions guaranteed before the next relabeling at that level is
  $\frac{\tau-1}{2}(2/\tau)^i$, not $(2 - \tau/2)(2/\tau)^i$ as stated: too large by a factor
  $(4-\tau)/(\tau-1)$, which is 5 at $\tau = 1.5$. A simulation confirms it: with $\tau = 1.5$ and
  insertions at the front, the fewest insertions between two relabelings at level $37$ were
  $33{,}707$, below the stated bound ($52{,}438$) and above the corrected one ($10{,}488$). The
  $O(\log n)$ amortized bound survives, with cost $2/(\tau-1)$ per level instead of $2 - \tau/2$, but
  the space/time trade-off of their Section~4.2 does not: the corrected bound $2\log
  n/((\tau-1)(1-\log \tau))$ first falls and then rises with $\tau$, with its minimum near $\tau = 1.46$.
  Their experiments count actual relabelings and are unaffected. The sentence in question
  reads: ``This requires at least $(2 - T/2)(2/T)^i$ insertions, thus yielding an amortized
  insertion cost of $2 - T/2 = O(1)$'' (proof of Lemma~3); it also calls a number of insertions
  a cost. The simulation is \texttt{experiments/bender-lemma3} in the repository.
\item \emph{Gaps.} Bender et al.\ defer their worst-case $O(1)$ algorithm to a full paper (Section~1),
  and leave a case of their file-maintenance algorithm and the proof of its Lemma~6 to the
  full version (Section~5); as of 2012 that version had not appeared~\cite{Kopelowitz12}.
  Kopelowitz's deterministic worst-case $O(1)$ construction (his Theorem~4.2) omits several
  details (Sections~4.1.2--4.1.5: ``details are omitted'', ``details are standard and are thus
  omitted''), and handles deletions only in a footnote, by marking and rebuilding, as we do;
  the bounds that hold in expectation in his paper are those of a generalization with subsets.
\end{itemize}

\paragraph*{In our own design.} Before the formalization we wrote a pen-and-paper proof. It
found one real bug and two gaps, all of which a testing campaign had missed (16 workloads,
about 5,300 random-parameter fuzz trials with every invariant checked after every operation,
and mutation testing). The campaign was then extended to 23 workloads, about 28,000 further
fuzz trials and a checker that measures every quantitative bound of the proof on every core.
\begin{itemize}
\item \emph{Rebuild sizing (a cost bug).} The prototype sized a new core by the number of live
  elements. But a rebuild lasts in proportion to the old core's size, tombstones included, and
  the user can insert throughout. An adversary fills a core to half, deletes every live element
  during the copy, and then inserts at the front: the next core, sized for no elements, receives
  every insertion while the copy crawls over the tombstones. At $2\cdot10^6$ elements its top
  table overflowed and single operations took $4{,}115$ prototype steps instead of about $100$. The fix
  sizes the new core for everything it can receive during the copy (\cref{sec:wrapper}).
\item \emph{Element label width (a gap).} The bound on insertions into one gap of a group
  allowed about $4G/3$ of them, but the label width paid for about $G$ halvings. Not observable
  at testable sizes; the width was increased.
\item \emph{Families split across chunks (a gap).} A chunk split could separate a splitting
  group from its siblings, so elements flowed into a chunk without insertions there, which
  breaks the accounting of the zeroing game. The fix moves whole families
  (\cref{sec:split}); as a side effect, two code paths became unreachable.
\end{itemize}
Completing the pen-and-paper proof and an independent adversarial review of each of its parts,
checked against the prototype, found further false steps, none of them in a final
conclusion: a lemma about group labels used an upper bound where it needed a lower one, two
claims about the nesting depth of label gaps were false as written (with counterexamples on the
real code), the rebuild inequality was off by one or two, and the rebuild needed a condition on
$\beta$ that was not stated (with $\beta = 3$ it produced equal tags). The formalization then
changed three parameters, not because a paper argument was false: $\beta$ and the group label
width because the mechanized arguments are coarser, and $K_R$ so that the waiting-chunks
inequality, which holds numerically with about 5\% to spare at $L = 8$ and 2\% at $L = 9$,
could be proved for every
$L \ge 8$ at once. $\beta$ went from 48 to 64, $K_R$ from 4 to 8, and the group label width
from about $9L$ to about $30L$ bits (\cref{app:params}).

\section{Transpilation and Evaluation}\label{app:eval}

\paragraph*{Transpiler and runtime.}
A Lean program of 186 lines transpiles the machine program to Rust statement for statement: a
machine local becomes a Rust variable, an array an index into the runtime's memory, a
\lean{call} a Rust function call, and \lean{fail} an error value propagated to the nearest
\lean{catch}. Every statement adds the cost that \lean{exec} charges for it to a counter, so
the generated code counts machine steps exactly. The runtime implements words as
arbitrary-precision naturals and provides the constant-time zero-filled allocation that the
model assumes (\cref{sec:limits}), using lazily zeroed memory and releasing retired cores a
few blocks per operation. Neither is verified. Instead, a Lean executable running
\lean{exec} and the generated code run the same operation files and print, for every
operation, its status, step count, answer and the bit length of the largest value computed
so far. The outputs agree byte for byte on nine operation files (13{,}920 operations, 7.7 million
steps, with many rebuilds). The larger tests compare the generated code with the
specification rather than with Lean: it agrees on 3.3 million random operations, 1{,}100 fuzz
traces (4.5 million records) and the 23 stress workloads of the prototype (9.0 million
records). The generated code is byte-identical to a fresh transpilation of the current
program, and the logs of all these runs are in \texttt{rust/results}.

\paragraph*{Benchmarks.}
We compare with two baselines, written by us in Rust: the amortized $O(1)$ algorithm of
Bender et al.~\cite{BenderEtAl02} (groups of at most 64 elements, $\tau = 1.4$, 62-bit tags),
and an AVL tree that stores the sequence in order. The insertion workloads insert after a
random element, after the first element (clustered insertion, the classic adversary for
relabeling), at the end and at the front; \cref{tab:native} reports the first two, together
with workloads that mix deletions and queries. The machine is an Intel Core Ultra 9 285K
under Linux 7.0, one performance core per run. Every operation is timed, and each configuration runs alone. Each workload is deterministic
and runs three times. We report the mean, the 99.99th percentile of each run (the median over
the runs), and the slowest repeatable time: each operation's minimum over the runs, of which we
take the largest. Only the operation itself is timed; its arguments are chosen beforehand. The minimum removes pauses that hit an operation in only some runs, and so also recurring
costs that move between operations; the per-run percentile keeps them, which is why it can exceed
the slowest repeatable time. The means of configurations whose code did not change differ by
up to 5\% between sessions and seeds, and the slowest repeatable time by up to about 90\%. One
pass is excluded from the tables: the first warm-query pass of the final session, run directly
after the $10^8$ insertions, was 4\% (native) to 30\% (AVL) slower than the pass we report,
which changes the ratios between the implementations. A rerun matched the earlier sessions, so we report the rerun; both passes are kept with the
results. The excluded pass suggests that the machine's state after a large run
matters, which our procedure did not control for. Every driver gives the three implementations the
same bound on the number of insertions: the native implementation reserves its handle table
from it (a mapping that grows one page at a time), while its cores are sized from the current
number of elements and rebuilt as the sequence grows and shrinks; the AVL tree and the
amortized baseline size their arrays from it.

\paragraph*{The transpiled program.}
Both baselines are much faster on average: the program performs about 1{,}100--1{,}500 machine steps per
insertion on average, computes with arbitrary-precision words and counts every step. Its slowest
insertion grows from 0.12--0.14\,ms at $10^5$ elements to 0.29--0.37\,ms at $5 \cdot 10^7$.
The figures are similar on all four workloads up to $10^7$ elements and on the two workloads
run at $5 \cdot 10^7$. The amortized baseline's slowest insertion, by contrast, grows linearly
under clustered insertion, to about 5\,ms at $10^8$. In machine steps, the slowest insertion took
$1.3 \cdot 10^5$ at $10^5$ elements and $3.1 \cdot 10^5$ at $5 \cdot 10^7$, against the proved
bound of $1.1 \cdot 10^9$. It grows because, as the top tree gets taller, more of the fixed
number of job steps per operation start a drain; that number is bounded, so the maximum is
bounded too. An order query always takes 65 machine steps in all, wrapper included; the bound
of 80 in \lean{order\_agree} is not tight.

\paragraph*{The native implementation.}
The native implementation (\texttt{experiments/lean-design} in the repository) has the same
core, wrapper and parameters, but uses records, fixed-width labels (128-bit element labels and
1,024-bit group labels, wide enough for the proof's widths at every measured size) and no step
counting; it is not verified. Level F and the native implementation both reproduce 28
prototype traces exactly (handles, keys, query answers, rebuild counters and key digests):
14.6 million operations, with up to 804 rebuilds in one trace. Across sizes and workloads its
mean insertion time is 1.3--2.8 times the AVL tree's, its 99.99th percentile 0.8--1.0 times
and its slowest repeatable insertion about 1.3--2.3 times (\cref{tab:native}). Under churn and
teardown its mean times are 0.4--2.7 times the tree's. Its cold order queries are about 6.3 times
faster than the tree's after random insertions and 2.2--2.3 times faster after clustered ones,
by either estimate of the mean (\cref{tab:eval-summary}). It
needs 112 bytes per element, against 13 for the tree and 21 for the amortized structure,
partly because retired cores are kept for reuse.

\begin{table}[t]
\centering
\caption{The native implementation of our algorithm (not verified) against the baselines, in
$\mu$s: mean / 99.99th percentile of a run / slowest repeatable time (\cref{app:eval}, Benchmarks). Insertions: those that grow the sequence from $10^7$ to $10^8$ elements. Churn: $10^7$
rounds of one deletion and one insertion in a sequence of $10^7$ elements, at random positions, or
as a queue (insertion at the end, deletion at the front). Teardown: $10^7$ random deletions down
to the empty sequence. Warm queries: one query on a random pair after each insertion, at $10^6$--$10^7$ elements.
Cold queries: random pairs at $10^7$--$10^8$ elements after random insertions, with cold caches,
measured and aggregated as in \cref{tab:eval-summary}. Memory: peak resident bytes per element after $10^8$
random insertions / peak over churn and teardown.}
\label{tab:native}
\small
\begin{tabular}{lrrr}
\toprule
 & native & AVL & amortized \\
\midrule
random insertion & 1.00 / 5.3 / 12.1 & 0.77 / 5.5 / 7.2 & 0.54 / 12.3 / 15.3 \\
clustered insertion & 0.085 / 2.9 / 6.3 & 0.030 / 3.2 / 3.6 & 0.022 / 3.4 / 5{,}050 \\
random churn: deletion & 0.29 / 3.7 / 11.0 & 0.71 / 4.0 / 7.0 & 0.19 / 1.2 / 0.39 \\
\phantom{random churn:} insertion & 0.73 / 4.2 / 5.3 & 0.68 / 5.4 / 6.0 & 0.33 / 9.8 / 12.1 \\
queue churn: deletion & 0.051 / 2.5 / 3.9 & 0.019 / 0.47 / 0.83 & 0.011 / 0.15 / 0.30 \\
\phantom{queue churn:} insertion & 0.053 / 2.3 / 5.6 & 0.027 / 3.4 / 3.4 & 0.022 / 3.7 / 557 \\
teardown: deletion & 0.39 / 3.1 / 7.8 & 0.73 / 4.2 / 5.9 & 0.19 / 1.3 / 0.76 \\
order query, warm & 0.44 / 1.6 / 0.76 & 0.40 / 1.9 / 1.4 & 0.13 / 1.1 / 0.37 \\
order query, cold & 0.93 / -- / 1.4 & 5.9 / -- / 8.4 & 0.50 / -- / 0.88 \\
memory (bytes per element) & 112 / 250 & 13 / 26 & 21 / 42 \\
\bottomrule
\end{tabular}
\end{table}

\section{Earlier Worst-Case Constructions}\label{sec:bg-prior}

Dietz and Sleator~\cite{DietzSleator87,DietzSleator88} gave the first worst-case $O(1)$
algorithm, and, as far as we know, the only one with published pseudocode (in the appendix of the technical
report); the pseudocode treats Willard's algorithm as a black box and omits the time-stamping
of its labels. Their structure is a tree of height four whose leaves are the elements; an
internal node of height $h$ has at most $\log^h N$ leaves, overfull nodes split
incrementally, and the children of the root are positions in Willard's file. Willard's
algorithm alone takes 54 pages, as Bender et al.\ point out~\cite{BenderEtAl02}. We
keep the overall shape of this construction (levels, incremental splitting and the zeroing
game) and replace Willard's algorithm.

Bender et al.~\cite{BenderEtAl02} gave a simpler amortized algorithm, relating list labeling
to weight-balanced trees, and announced a simpler worst-case $O(1)$ algorithm, obtained
by deamortizing it, for a full paper that, as of 2012, had not appeared~\cite{Kopelowitz12}. Our
top tree is weight-balanced too.

Kopelowitz~\cite{Kopelowitz12} gave a deterministic worst-case $O(1)$ construction, with some
details omitted, as the basis of a generalization with subsets whose bounds hold in expectation. Its
shape is close to ours: chunks of $O(\log^2 n)$ elements, split largest-first, are labeled by a
weight-balanced B-tree~\cite{ArgeVitter03} instead of Willard's algorithm, and each insertion
into the tree is spread over the following $\log n$ insertions. His tree relabels in place, in
the background, with an extra most significant bit that keeps the order during a rewrite.

In list-labeling terms, both top trees label the chunks from a universe polynomial in $N$.
Our tags have $H\beta = 64(\lfloor L/3\rfloor + 1)$ bits, so the universe has roughly
$N^{21}$ labels. This leaves room for a separate label field per level of the tree: draining a
node changes only two fields in the tag of each chunk below it, and a weight-balanced node of weight $W$
gains $\Omega(W)$ chunks before it splits again~\cite{ArgeVitter03}, so the rewrite can be
paid in constant instalments. Willard's algorithm solves file maintenance, whose universe is
linear in the number of items and leaves no such room. The price is width: about $21L$ bits
per tag instead of $L + O(1)$, still a constant number of words.

Bender et al.~\cite{BenderEtAl02} had already sketched a simpler worst-case file-maintenance
algorithm, leaving a case and a proof to the full version (\cref{sec:errors}). Bender et
al.~\cite{BenderEtAl17} later gave a simpler deamortized algorithm for sequential file
maintenance, which could take the place of Willard's at the top of Dietz and
Sleator's construction. Like Willard's, it moves elements, that is, it relabels in place; our
top tree avoids this with immutable tables.

\section{Related Techniques and Verified Complexity}\label{app:related}

\paragraph*{Order maintenance and list labeling.} We build on Dietz and Sleator's incremental splitting and zeroing game
\cite{DietzSleator87,DietzSleator88}, on the labeling perspective of Bender
et al.~\cite{BenderEtAl02}, and on Kopelowitz's weight-balanced top tree~\cite{Kopelowitz12}.
Order maintenance is closely related to online list labeling,
where labels come from a fixed range: with a range linear in $n$, $\Theta(\log^2 n)$ amortized
relabelings per insertion are sufficient~\cite{ItaiEtAl81} and, for deterministic
algorithms, necessary~\cite{BulanekEtAl12}, and randomized algorithms do better~\cite{BenderEtAl22}. Bender et al.\ compared amortized
order-maintenance structures experimentally~\cite{BenderEtAl02}; our top tree is a
weight-balanced tree in the sense used by Arge and Vitter~\cite{ArgeVitter03}. As far as we
know, no worst-case $O(1)$ order-maintenance algorithm has been implemented before, and no
order-maintenance algorithm has been verified in a proof assistant.

\paragraph*{Deamortization.} Our algorithm turns every expensive step into background work
spread over later operations: splits move a few members per insertion, top jobs advance a
few steps per insertion, and the structure itself is replaced by global
rebuilding~\cite{Overmars83}. The same idea underlies real-time functional data structures,
where scheduled lazy evaluation pays for expensive operations in
advance~\cite{Okasaki98}. Real-time deques are the closest verified relatives of our
structure. Tóth and Nipkow verify the correctness of Chuang and Goldberg's real-time
double-ended queue in Isabelle~\cite{TothNipkow23}, and Viennot et al.\ that of Kaplan and
Tarjan's catenable deques in Rocq~\cite{ViennotEtAl25}. Tóth and Nipkow argue the
worst-case $O(1)$ bound informally, from the absence of recursion. Viennot et al.\ make that
argument for a separate, hand-written OCaml implementation; their verified Rocq code keeps
index computations, so it does not attain the constant bound, and the OCaml code is not
formally related to it. The Hood--Melville
queue of the Archive of Formal Proofs is likewise verified for correctness
only~\cite{GomezLondono21}. Madhavan et al.\ prove concrete worst-case step bounds for
Okasaki's real-time queue and deque with an SMT-based verifier, against an instrumented
semantics whose soundness is proved on paper~\cite{MadhavanEtAl17}; their bounds rest on
invariants of the memoization cache and the schedule.

\paragraph*{Verified data structures.} The correctness of many functional search trees has
been verified, largely automatically, in Isabelle/HOL~\cite{Nipkow16}, and a textbook develops
correctness and running-time proofs for a broad collection of functional data
structures~\cite{NipkowEtAl25}. Charguéraud and Pottier verify an
imperative union-find structure, with its amortized complexity, in separation
logic~\cite{ChargueraudPottier19}. Our cost proof must maintain interacting size and scheduling invariants by induction over time (\cref{sec:why}).

\paragraph*{Verified complexity.} Most mechanized complexity results are amortized or
asymptotic, and most attach costs to a program logic for a real language. Time credits in
separation logic, introduced by Atkey for amortized resource analysis~\cite{Atkey10}, are used
in CFML for amortized~\cite{ChargueraudPottier15,ChargueraudPottier19} and
asymptotic~\cite{GueneauEtAl18} bounds, including those of an incremental cycle detection
algorithm~\cite{GueneauEtAl19}, and time receipts in Iris bound counters by the elapsed
time~\cite{MevelEtAl19}.
In Isabelle/HOL, Nipkow and Brinkop verify the amortized complexity of self-adjusting data
structures~\cite{NipkowBrinkop19}, Haslbeck and Nipkow study Hoare logics for time
bounds~\cite{HaslbeckNipkow18}, Zhan and Haslbeck verify asymptotic bounds of imperative
programs~\cite{ZhanHaslbeck18}, and Haslbeck and Lammich refine running times from abstract
algorithms to imperative code~\cite{HaslbeckLammich19} and down to LLVM~\cite{HaslbeckLammich21}. In type theory, Danielsson
analyses lazy functional data structures with a cost monad~\cite{Danielsson08}, and
\textsf{calf} is a framework for cost-aware programs~\cite{NiuEtAl22}. Automatic amortized
resource analysis infers bounds for OCaml programs in RAML~\cite{HoffmannEtAl17}, and Liquid
Haskell checks resource bounds of programs annotated with explicit
ticks~\cite{HandleyEtAl20}. Our setting differs from most of these in two ways. The bound is a
concrete worst-case constant, as in Madhavan et al.'s work~\cite{MadhavanEtAl17}, not an
amortized or asymptotic one, and it rests on a timing argument across operations about
invariants of the whole structure rather than on the shape of the code. And the cost comes
from the interpreter of a deeply embedded language, as in Haslbeck and Nipkow's Hoare logics
for time~\cite{HaslbeckNipkow18}, rather than from a cost monad with explicit ticks, between
which uncounted work is possible. Time credits also rule out uncounted work, charged at every
function call rather than at every instruction. The price of our choice is a compiler from a
structured language to the machine. CerCo's cost-annotating compiler~\cite{AmadioEtAl13}
compiles C to assembly and must account for the cost of each source fragment; ours maps between
two languages designed so that step counts correspond exactly, which makes that
correspondence provable. Quantitative CompCert preserves stack-space bounds through a
realistic compiler~\cite{CarbonneauxEtAl14}, and Carbonneaux et al.\ infer certified resource
bounds for C automatically, by amortized analysis with potentials~\cite{CarbonneauxEtAl15};
our bound needs coupled structural and scheduling invariants beyond a local amortized
analysis. Pottier et al.\
reconstruct Okasaki's debits with time credits~\cite{PottierEtAl24}; the bounds remain
amortized.

\paragraph*{Value sizes and machine models.} Theorem W answers a question that verified cost
models usually settle in other ways. Forster, Kunze and Roth show that the weak call-by-value
$\lambda$-calculus, with $\beta$-steps as time and term size as space, is a reasonable machine
model~\cite{ForsterEtAl20}, and mechanizations of the Cook--Levin theorem either build on such an
invariance result~\cite{GaeherKunze21} or work with Turing machines directly~\cite{Balbach23}. They settle the question once for a model;
we prove a bound on value sizes for one program on a unit-cost machine. CakeML's space cost
semantics is the closest analogue: it proves at the source level that a program's heap and
stack fit the machine, so that compiler correctness holds without the out-of-memory
exception~\cite{GomezLondonoEtAl20}. Haslbeck and Lammich implement natural numbers by 64-bit
words and exclude overflow by size preconditions that propagate to the program's
precondition~\cite{HaslbeckLammich22}; we instead let the word size grow with the number of
operations and prove how large it must be. Deductive verifiers prove the absence of overflow
operation by operation, as SPARK does for run-time errors~\cite{McCormickChapin15}. Clochard et
al.\ avoid such proofs for counters that grow by one per step, which cannot overflow in feasible
time~\cite{ClochardEtAl15}; time receipts mechanize the same argument~\cite{MevelEtAl19}, and it
is the one that applies to our handles.

\end{document}